\documentclass[12pt,preprint]{aastex}
\slugcomment{Accepted and scheduled for publication  
in {\it The Astrophysical Journal},  for  the February  2012, v 745, 1st issue}  
\def\lax {\ifmmode{_<\atop^{\sim}}\else{${_<\atop^{\sim}}$}\fi}  
\def\gax {\ifmmode{_>\atop^{\sim}}\else{${_>\atop^{\sim}}$}\fi}  
\def\gtorder{\mathrel{\raise.3ex\hbox{$>$}\mkern-14mu
             \lower0.6ex\hbox{$\sim$}}}

\def\cm2{cm$^{-2}$}
\def\s1{s$^{-1}$}

\begin{document}

\title{GX~3+1: the stability of spectral index as a function of mass accretion rate}

%\title{On the Constancy of the Photon Index of  X-ray spectra of 4U~1728-34 through all spectral states} 
%during outburst transitions}

\author{ Elena Seifina\altaffilmark{1} and Lev Titarchuk\altaffilmark{2}}
\altaffiltext{1}{Moscow State University/Sternberg Astronomical Institute, Universitetsky 
Prospect 13, Moscow, 119992, Russia; seif@sai.msu.ru}
\altaffiltext{2}{Dipartimento di Fisica, Universit\`a di Ferrara, Via Saragat 1, I-44100 Ferrara, Italy, email:titarchuk@fe.infn.it; 
%ICRANET, Piazza della Repubblica 10-12 65122 Pescara,  Italy; 
George Mason University,  School of Physics, Astronomy and Computational Sciences (SPACS), Fairfax , VA 22030 ltitarch@gmu.edu;   
Goddard Space Flight Center, NASA,  code 663, Greenbelt  
MD 20770, USA; email:lev@milkyway.gsfc.nasa.gov, USA}

\begin{abstract}
We present an analysis of the spectral and timing properties observed in X-rays from 
neutron star  binary GX~3+1 (4U~1744-26) during long-term transitions between 
the {\it faint} and 
{\it bright} phases superimposed on short-term %$mild$ (lockal, short on time scale of hours) 
transitions %(passages, movings) 
between %from 
{\it lower banana} (LB) 
and %to 
{\it upper banana} (UB) 
branches in terms of its color-color diagram.   
%when electron temperature $kT_e$ of the Compton cloud changes from 4.5 keV to 2.3 keV. 
We analyze all observations % number of  transition episodes 
of this source  obtained with the 
{\it Rossi} X-ray Timing Explorer
%({\it RXTE}) 
and {\it Beppo}SAX satellites.   We find that the X-ray broad-band energy 
spectra during  these spectral  transitions %spectral states 
can be adequately reproduced by  a composition of  a
low-temperature blackbody component, a Comptonized component ({\it COMPTB}) and  {\it Gaussian} 
component. 
%%% BEGIN
%{\it
We argue %This model allow to argue %Spectral analysis using this model provides 
that the electron temperature $kT_e$ of the Compton cloud 
monotonically increases from 2.3 keV to 4.5 keV, when GX~3+1 makes a transition  from  UB to  LB. 
%related to  low flux variabilities on time scale of hours.  
%Specifically, we
We also detect %demonstrate 
an evolution of noise components (a very low frequency noise and a high frequency noise) 
%that 
during these LB -- UB transitions. %, associated with low flux variability on time of hours, 
%the power spectra of X-ray signal in {\it upper banana} are dominated by a very low frequency noise (VLFN) with 
%%peaked noise component 
%the breake at around 20 Hz, whereas in {\it lower banana} the power spectrum are dominated by a high frequency 
%noise (HFN) in 1 -- 100 Hz range and acompanied by reduced VLFN below $\sim$ 1 Hz.
%}
%%%%% END
Using  a disk {\it seed} photon normalization of {\it COMPTB}, which is 
proportional to the  
%(sub-Keplerian plus  disk) 
mass accretion rate, we find that the 
photon power-law index $\Gamma$  is almost constant ($\Gamma=2.00\pm 0.02$) when 
%$kT_e$ changes from 2.3 to 3.7 keV 
mass accretion rate  changes by factor 4.
%from 0.04 to 0.13 $L_{39}/D_{10}^2$ erg/s/kpc$^2$
%during the  spectral transitions.
In addition, we find that the emergent spectrum is  dominated 
by the strong 
%thermal 
Comptonized component. 
We interpret  this quasi-stability of the index $\Gamma$ and a particular form of the spectrum
in the framework of a model in which the energy release in the transition layer  located between the  
accretion disk and neutron star surface dominates  that in the disk. 
% In this case   the spectrum  ormed  
%The index quasi-stability 
%takes place when the  energy release in the TL is much higher than the flux  coming to the TL 
%from the accretion disk. 
Moreover, this index stability effect now established for  GX~3+1 
%during state %spectral 
%evolution of the source 
%from the low to high luminosity states 
was previously found in  the  $atoll$ source 4U~1728-34 and  suggested for a number  of other low mass X-ray neutron star 
%LMXB NS 
binaries  (see Farinelli \& Titarchuk).
 This   intrinsic behavior of  neutron stars, in particular for $atoll$ sources,  is fundamentally  
different   from that  seen in black hole binary sources where the index monotonically increases during 
spectral transition from the low state  to the high state and then finally   saturates at high values of mass accretion rate.

\end{abstract}


\section{Introduction}
Low mass X-ray binaries (LMXB) hosting a neutron star (NS) show a variety of spectral states and transitions between them. 
%$some$ of them %are transient sources that  
%demonstrate transitions between different states. 
%significant flux variability in a wide range of time scales. 
In this regard, the so-called {\it atoll}  %class 
sources [see e.g. 
\cite{hasinger89}, \cite{klis05}]
%Hasinger \& van der Klis 1989, van der Klis 2005) are particularly 
are particularly interesting because they demonstrate a wide range of luminosities [the most of them show 
from 0.01  to  0.5 of  the Eddington limit $L_{Edd}$].
% where \sim 2\times 10^{38}$ erg s$^{-1}$ depending on helium abundance]. 
It is worth noting  that the name of {\it atoll} sources  is  associated  with the shape traced in the color-color diagram 
(CD). This shape  can be divided into  two main 
regions, corresponding to two X-ray states of the source: the harder one is related  to the island (IS) state  and the softer one is related to  the banana (B) state.
%%successive transitions from  
%%the usually  called  
%%{\it island}  state (IS)  through the intermediate ({\it lower \& lower left banana}) state (LLB, LB)  
%%to the {\it upper banana} state (UB) on time scale of hours. During long time scale these objects sometimes 
%% demonsrate  long-term transitions between the {\it low luminosity} (LS) and {\it high luminosity) (HS) states.
%low ({\it island})  state (LS)  through the intermediate ({\it banana}) state (IS)  
%to the high ({\it upper banana}) state (HS). 
%The so-called {\it atoll}  class sources [see e.g. 
%\cite{hasinger89}, \cite{klis05}]
%%Hasinger \& van der Klis 1989, van der Klis 2005) are particularly 
%are particularly interesting because they  occupy  
%Further, 

%During the  transitions between these  states  
%atoll sources %usiallygh demonstrate %show 
%a wide variety  of spectral states 
%are characterized by  different spectral 
%shapes.  
These spectra of NS sources can be described by     {\it blackbody}  (BB) components,  with color temperatures  $kT_{BB}<1$ keV and $kT_s>1$ keV that are  presumably  
related to the accretion  disk and NS surface respectively. In addition, there is  a   thermal  Comptonization component with   electron temperature $kT_e=2.3-15$ keV that is  probably 
related  to  the  transition layer (TL)  located between the disk and  NS 
%characterized by  electron plasma temperature 
[see \cite{P06}; \cite{ft11}; \cite{st11}, hereafter FT11 and ST11 respectively]. 
An analysis of X-ray power spectra of atoll sources indicates to a tight relation between timing properties and the position 
on the  %color-color diagram 
CD, suggesting that the source timing and spectral properties are well determined 
by basic parameter such as mass accretion rate 
%and have a common origin in physical processes in the inner part of the  accretion flow 
[see, e.g., \cite{disalvo2001}].

%%BEGIN

GX~3+1 is one of the brightest atoll sources associated with a bulge component of our Galaxy. 
GX~3+1  along  
%together 
with GX~9+9, GX~9+1 and GX~13+1 form the subclass of persistently bright atoll sources,
%(or bright atoll sources)
 which are always  in the 
banana state [see \cite{hasinger89}].
  In particular,   two-branch structures have been observed
in the CD and hardness-intensity diagram (HID) of GX~3+1 [\cite{stella85}; %Stella et al. 1985; 
\citet{Lew87}; %Lewin et al. 1987; 
\citet{schulz89}; \cite{hom98}; %Homan et al. 1998; 
\citet{Muno02}; %Muno et al. 2002; 
\citet{Schnerr03}]. %Schnerr et al. 2003]. 
Specifically, their tracks in the X-ray CD %color-color diagram
are long, diagonal and slightly curved, while their fast timing properties are %always (usually) 
dominated only by a relatively weak (1 -- 4\% rms) power-law shaped noise component.  These aforementioned  atoll  sources are intermediate in terms of luminosity  that changes in the range 
0.1-0.5 of $L_{Edd}$ [see  \cite{chsw97} and  \cite{Ford00}].
%between the very luminous
%Z-sources and the weaker remaining atoll sources 
%Ford et al. 2000), 
%with a luminosity  %times the Eddington limit 
%[see]. %Christian \& Swank 1997). %, with $L_{Edd}=2.5\times 10^{38}$ erg s$^{-1}$). 
%At low luminosity states GX~3+1 displays the properties of so-called ``more frequently bursting atoll sources'' 
%and demonstrate Type~I X-ray bursts providing a neutron star nature for compact object of this binary.

%{\it  
In contrast %Contrary 
to  other atoll sources and  Z sources,  these bright atoll sources have so far  not shown  kHz
QPOs [see \cite{Wijn98}; %Wijnands et al. 1998b; 
\cite{stroh98}; %Strohmayer 1998; 
\cite{hom98}; %Homan et al.  1998; 
\cite{ooster01}; %Oosterbroek et al. 2000; 
\cite{Schnerr03}], % Schnerr et al. 2003, 
which can be attributed  to the fact  that  these objects do not  reach the {\it left lower banana} (LLB), where other atoll  sources exhibit kHz QPOs [\citet{vdKl00}]. %van der Klis 2000). 
For example, the  weaker atoll sources, such as e.g. 4U~1608-52, 4U~0614+09 
or  4U~1728-34, show kHz QPOs in LLB. % with typical rms values of about 3 -- 15\% and with a significance of 4 - 20.

 X-ray data of  GX~3+1 (4U~1744-26) obtained  in broad  energy ranges using   $Beppo$SAX (0.1 -- 200 keV) and {\it RXTE} (3 -- 200 keV)  offer 
%and intensive broad band studies, mainly 
%performed with instruments  open 
a unique possibility to further investigate, in detail,  the evolution of X-ray spectral properties during 
transition %spectral state 
events. This bright  {\it atoll} source shows long-term transitions 
from the {\it fainter phase} %{low luminosity} soft state (LS) 
to its {\it brighter phase} in X-rays %high luminosity} soft state (HS)
%\footnote{
%Here the termins ``{\it low state}'' and ``{\it high state}'' use only for indicating of ``$fainter$'' and 
%``$brighter$'' intervals of GX~3+1 ligh curve (see Fig.~1) and differ from termins ``{\it low state}'' and 
%``{\it high state}'' commonly adopted for describing of spectral state evolution in X-ray binaries hosting 
%{\it black hole candidats}.
%}
 and vice versa when the corresponding  
luminosity changes, at least, by  a factor of  4, while  on time scales of hours GX~3+1 demonstrates low flux variabilities 
as transitions between LB and UB states.  Naturally  one can pose a fair question: what is  the physical mechanism responsible for the source  emission during %quasistability
 these luminosity  changes and particularly  how  the spectral index changes during these  transitions?  %changes.
 % of the dependence of 
%index  vs  can be found.
% The bright Galactic bulge source 

GX~3+1  was discovered   during an $Aerobee$-rocket flight on June 16, 1964 \citep{bow65}.  Subsequently, this source   was observed  many times during   various observational campaigns.
%Bowyer et al. 1965).
Detailed long-term monitoring observations of GX~3+1 was performed by 
{\it All Sky Monitor}  on $GINGA$  [see \cite{asai93}]; EXOSAT [see \cite{schulz89}]; 
%Asai et al. 1993) and
 {\it RXTE} [see \cite{bradt93}; 
%Kuulkers \& van der Klis 2000]
\cite{kk00}],  %and also by 
{\it Wide Field Camera} (WFC) of $Beppo$SAX~[see \cite{denhart03}] 
and also by INTEGRAL [see \cite{P06}].
In particular,  \cite{denhart03} found  
three types  of variability: $short$-$term$ variation (of order of seconds), 
 $mild$ variability on a daily (hourly) time scale and  $slow$  sinusoidal-like variation on a time scale 
of years.
% [see \cite{mak83}].
%Makishima et al. 1983). 
However, it is surprising that the hardness ratio, 
which can be  a measure of the spectral shape, stays almost 
constant during these observations.  
%variabilities 

Although an optical counterpart has not yet been identified [e.g. \cite{n91}] 
%Naylor et al. 1991), 
GX~3+1 is presumably a
low mass X-ray binary in which a neutron star is accompanied by a low-mass star of spectral type A or
a later. During an active stage the companion overflows its Roche lobe and transfers matter
onto the NS via an accretion disk. This process is possibly accompanied by
nuclear burning of  helium or hydrogen  layer of NS surface  
as a  result of the matter accumulation on the NS surface \citep{hh75}.
%Hansen \& van Horn 1975).
Unstable fusion occurs leading to  thermonuclear flashes which can be observable in the form of 
X-ray bursts, so-called type-I X-ray bursts  \citep{wt76}. 
%and \cite{mc77}].
%Woosley \& Taam 1976; Maraschi \& Cavaliere 1977). 
GX~3+1 does exhibit fast variability in the form of  type-I X-ray bursts, which were 
extensively studied by a number  of X-ray missions $GINGA$ \citep{asai93}, 
%Asai et al. 1993), 
$Granat$ [\cite{p94}, \cite{molkov99}]
%}Pavlinsky et al. 1994) 
and  by {\it RXTE} [\cite{kk00} and  \cite{k02}]  and by INTEGRAL \citep{ch06}. 
% Kuulkers 2002), and INTEGRAL 
%(Chenevez et al., 2006). 

A unique super-burst with a 
decay time of 1.6 hours 
%(Kuulkers 2002) 
was detected with the All Sky Monitor (ASM) on {\it RXTE} \citep{k02}.
One of the shortest bursts ever seen  exhibits a quick (i.e., less than 2 seconds) radius expansion 
%(factor of two)
phase, indicating that the burst luminosity was at the Eddington luminosity, causing the NS atmosphere
to expand due to radiation pressure. This implies a distance to the source is about 4.5 kpc,  assuming that  the NS atmosphere is  
hydrogen-rich [see more details in \cite{kk00}].
%(Kuulkers \& van der Klis 2000).

In this Paper we  concentrate our efforts  on the spectral and timing  properties of GX~3+1 related to changes in
mass accretion rate, which are seen  as the $mild$ and $slow$ variabilities. Previously,  an analysis of the burst 
properties of GX~3+1 as a function of mass accretion rate on the time scales larger than one minute  were presented 
by \cite{asai93} and \cite{denhart03}. In particular, the $slow$ variability has been revealed during transitions from
the {\it faint phase} %so-called ``low luminosity'' 
to the  {\it bright phase} of luminosity %``high luminosity'' states %(Fig. 1)
and are generally caused by significant increase of mass accretion rate.   The $mild$ variability is presumably  
related to moderate  local variability of mass accretion rate and revealed as local transitions between 
{\it lower banana} and {\it upper banana} states. The $slow$ variability has been investigated with the {\it RXTE}/ASM 
light curve, {\it Wide Field Camera} (WFC) of  $Beppo$SAX \citep{denhart03} and 
%(den Hartog et al., 2003) and  
the all sky monitor on $GINGA$ \citep{asai93} and these observations indicate  that the flux oscillates semi-sinusoidally with a period of 6 -- 7 years (see Fig. \ref{variability_96-10}).  

 Moreover, the $GINGA$ measurements with its Large Area Detector  \citep{asai93} show a constant 1 -- 20 keV spectral 
shape over the {\it  fainter}  phases %``low luminosity''  state 
and  also during  {\it brighter} phases on luminosity %``high luminosity'' state 
%(Asai et al. 1993), 
  suggest a constancy of the spectral index.
  % during all spectral state transitions. 
This stability of the X-ray spectral shape over the {\it bright} %``low'' 
and {faint} %``low'' 
phases %``high'' states 
was  also been confirmed by the WFC {\it Beppo}SAX measurements in the 2 -- 28 keV band which showed minimal  spectral changes \citep{denhart03}.
%changes are minimal (den Hartog et al., 2003).
% with gradient (drop) of the electron temperature $kT_e$ of Compton cloud. 
%(den Hartog et al., 2003; Asai et al., 1993) exhibits....

The stability of the index was noted previously by  FT11,  
for a number of NS LMXB at different luminosities. They collected  X-ray spectra obtained by {\it Beppo}SAX and 
demonstrated the relative stability of spectral index $\alpha$ approximately of 1 ($\Gamma=\alpha+1$) for 
quite a few NS sources X~1658-298, GX~354-0, GS~1826-238, 1E~1724-3045, Cyg~X-1, Sco~X-1, GX~17+2 and 
%GX~3+1,
 GX~349+2
at different spectral states. Recently ST11 presented results of analysis of X-ray spectra for  {\it atoll} source 4U~1728-34,  detected with {\it Beppo}SAX and {\it RTXE} at different luminosities and values of the best-fit electron temperatures.  These  results  indicate   that  the value of  the index  varies slightly  about a value of 1, (or the photon index varies around 2)  independently of the electron temperature of the Compton cloud $kT_e$ and luminosity.  
%at least for this particular sample (``atoll'' class) of NS spectra [see \citet{disalvo2000a}; \citet{F08}].
 This unique stability of the photon index may be  an intrinsic 
property of NS at least for the  {\it atoll}  sources.  
It  is probably  determined  by common physical conditions for  this class of sources.  FT11 \& ST11  interpreted  this quasi-stability of the index $\Gamma$ 
in the framework  of a model in which the spectrum  is  dominated by a strong thermal Comptonized component formed  
in the transition layer (TL)  located between the accretion disk and neutron star surface. 
Indeed, the index quasi-stability takes place when the  energy release in the TL is much higher than  the photon energy flux   coming from the accretion disk  and illuminating  the TL.

The evolution of spectral parameters of compact objects in X-ray binaries is 
of great interest for understanding their nature.
% of compact objects. 
It is well known that many  black hole (BH) candidate binaries exhibit  correlations 
between mass accretion rate $\dot M$ and photon power-law index $\Gamma$  
%during  spectral transition when sources evolve from the hard  to soft states 
[see  \citet{st09} and \citet{tsei09}, hereafter ST09 and TS09, respectively].  In the soft states of BHs 
these index-$\dot M$ correlations almost in any case  show a saturation  of $\Gamma$  at high values of the 
$\dot M$.  This saturation effect can be  considered   
as a black hole signature or equivalently as a signature of a converging flow into BH (ST09 and TS09).

In this Paper we present the analysis of the {\it Beppo}SAX  available observations 
%during  1998 -- 1999 years
 and  {\it RXTE}/PCA observations during  1996 -- 2010 years for 
GX~3+1.  In \S 2 we present the list of observations used in our data analysis while 
in \S 3 we provide the details of X-ray spectral analysis.  We analyze an evolution of 
X-ray spectral and timing  properties during the state transition in \S 4.  
We  make our  conclusions in  \S 5.

\section{Data Selection \label{data}}

Broad band energy spectra of the source were obtained
combining data from  three {\it Beppo}SAX Narrow
Field Instruments (NFIs): the Low Energy Concentrator
Spectrometer (LECS) for 0.3 -- 4 keV  \citep{parmar97}, the Medium Energy Concentrator Spectrometer
(MECS) for 1.8 -- 10 keV \citep{boel97} and the Phoswich Detection
System (PhDS) for 15 -- 60 keV \citep{fron97}. 
 The SAXDAS data analysis package is used for processing data. 
For each of the instruments we performed the spectral analysis in the energy range 
for which response matrix is well determined. 
The LECS data have been 
re-normalized based on MECS. Relative normalization of the NFIs were treated 
as free parameters in
 model fitting, except for the MECS normalization that was fixed at a value
 of 1.  We then  checked 
 this  fitting procedure 
 if these normalizations
 were in a standard range for each
 instruments% (section 4.2 of Cookbook for the BeppoSAX NFI spectral analysis
\footnote{http://heasarc.nasa.gov/docs/sax/abc/saxabc/saxabc.html}.
% Specifically, LECS/MECS re-normalization ratio is 0.92 and PDS/MECS
% re-normalization ratio is 0.97. 
In addition,  spectra are rebinned  accordingly to  
 energy resolution of the instruments in order to obtain
 % independent 
significant data points.  We rebinned  the LECS spectra with a binning factor 
which is not constant over energy
(Sect.3.1.6 of Cookbook for the {\it Beppo}SAX NFI spectral analysis) using
 template files
 in GRPPHA of
 XSPEC \footnote{http://heasarc.gsfc.nasa.gov/FTP/sax/cal/responses/grouping}. Also we rebinned the PhDS spectra with a linear binning
 factor of  2, grouping two bins together (resulting bin width is 1 keV).  Systematic error of 1\% 
have been applied to these analyzed spectra. 
In Table 1 we  listed  the {\it Beppo}SAX observations used in our analysis. 

\begin{figure}[ptbptbptb]
\includegraphics[scale=0.9,angle=0]{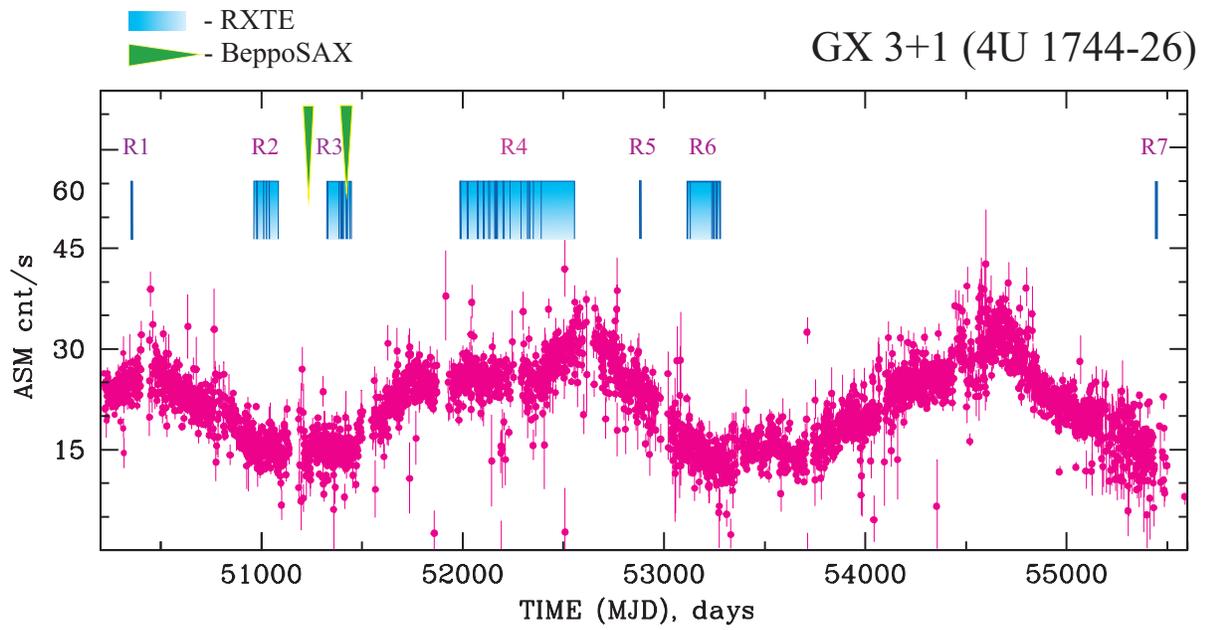}
\caption{  
 Evolution of ASM/{\it RXTE} count rate %, flux density $S_{8.46 GHz}$ at 8.46 GHz (VLA), 
%BMC normalization and 
%photon index $\Gamma$ 
during  1996 -- 2010 observations of GX~3+1. 
{\it Blue} vertical strips ({\it on the top of the panel}) indicate to time  for the {\it RXTE} 
pointed observations. 
%used in our
%analysis, 
Whereas
{\it bright blue} rectangles are related to  the {\it RXTE} data sets listed in Table 1, 
 {\it green} triangles show {\it Beppo}SAX NFI data, listed in Table 2.
%Blue rectangles indicate the {\it RXTE} data of pointed observations and green triangles show {\it Beppo}SAX NFI data.
%Black dashed line illustrates  mean count rate of 
%of  Dwell type ASM light curve and indicates long-term quasi-periodic variability of mean soft flux 
%during  $\sim$ six years cycle.
%b. Model flux in 3-10 keV and 10-50 keV energy ranges (blue and crimson points respectively).
%c. Electron temperature $kT_e$ in keV.
%d. {\it COMPTB} and blackbody normalizations (crimson and blue points respectively). 
%e. In the last bottom panel we present variations of spectral index $\alpha=\Gamma-1$.
% data of pointed observations and  
%green triangles show BeppoSAX NFI data used for the analysis.
}
\label{variability_96-10}
\end{figure}

We have   analyzed the  available data obtained with {\it RXTE}~\citep{bradt93} which have been 
%(Bradt et. al, 1993) 
found  in the  time period from October 1996 to September 2010 [see also  a review by  \citet{gall08}]. 
%In general, there are 106 bursts from
%this source in the {\it RXTE} burst catalogue \citet{gall08}. 
In our investigation we selected 
101 %99 
observations made at different count rates (luminosity states)
%}
% states 
%{\it (luminosity) }states of 4U~1728-34 
with a good coverage of rise-decay transition tracks. 
%%In general we investigate an evolution of X-ray emission from 4U 1728-34 during 50128-51733 MJD interval 
% {piraino00} 
%%within 
%one continuos  cycle of mean ASM flux variability (see Fig. \ref{outburst_05_rise}).
%{\it
We have  made an analysis of {\it RXTE} observations  of GX~3+1  during fourteen years 
for 7 intervals indicated by  blue rectangles in Figure~\ref{variability_96-10} ($top$).

%%%We have also  analyzed two {\it Beppo}SAX observations which dates  are shown in Table 1.
% by green triangles there.
%Figure~\ref{outburst_05_rise}.  
%and   \ref{evolution_lc_3}).
%}
 
{\it RXTE}/PCA spectra have been extracted and analyzed, wherein  PCA {\it Standard 2} mode data, collected 
in the 3 -- 50~keV energy range, using the most recent release of PCA response calibration (ftool pcarmf v11.1). 
The relevant deadtime corrections to energy spectra %and timing spectra 
have been applied.
% as following:
% The standard dead time correction procedure 
%has been applied to the data. 
%The average dead time  correction is in the range  3 -- 10 \% depending on the count rate value.
We used the data which  are available through the GSFC public archive 
(http://heasarc.gsfc.nasa.gov). 
In Table 2 we presented the groups  of {\it RXTE} observations which cover  
 the source evolution  from  {\it faint } %low} 
 to  {\it bright} %high} 
 (phase) events. Note that available {\it RXTE} data %set 
contains %(includes) 
one ``$bright$ phase'' set ($R4$) 
 and six ``$faint$ phase'' set ($R1$ -- $R3$, $R5$ -- $R7$). 
The PCA energy spectra were modeled using XSPEC astrophysical fitting software. 
%%%Spectral analysis was done using an approach similar to that adopted in ST11 for 4U~1728-34 data. 
Systematic error of 0.5\% 
have been applied to the analyzed spectra. 
 
We have also used public GX~3+1 data from the  All-Sky Monitor (ASM) 
on-board \textit{RXTE}, which show long-term quasi-periodic variability of the mean soft flux 
during  $\sim$ six years cycle (Fig.~\ref{variability_96-10}). 
%In the following 
We  use definitions of the  $fainter$ %low 
and $brighter$ %high 
on luminosity phases %states 
to relate these phases %states 
to the source luminosity and we demonstrate that during the bright/faint phase %low-high state 
transition of GX~3+1  COMPTB normalization
%the electron temperature of Compton cloud 
changes from 
%2.3 keV to 15 keV 
0.04 to 0.14 $L_{39}^{soft}/{D^2}_{10}$
% erg/s/kpc$^2$ 
where  $L_{39}^{soft}$ is the soft photon luminosity in units of $10^{39}$ erg/s  and  $D_{10}$ in units of  10 kpc  is distance to the source.

\section{Spectral Analysis \label{spectral analysis}}
In our spectral  data analysis we use a model which consists a  sum of a  Comptonization
  component ($COMPTB$) 
 [{\it COMPTB} is an XSPEC contributed model\footnote{http://heasarc.gsfc.nasa.gov/docs/software/lheasoft/xanadu/xspec/models/comptb.html},
see \citet{F08}, hereafter F08],  soft blackbody component of temperature 
 $T_{BB}$ and Gaussian line component.
  The $COMPTB$ spectral component has the following parameters:
temperature of the seed photons $T_s$,
energy index of the Comptonization spectrum $\alpha$ ($=\Gamma-1$), 
electron temperature $T_e$,   illumination (Comptonization) fraction $f$ of the Compton cloud by the soft (NS) photons, $f=A/(1+A)$, %which is the relative weight of the Comptonization component] 
and the normalization of the seed (NS) photon spectrum $N_{COM}$. 
%%%% BEGIN

We   include   a simple {\it Gaussian} component in the model, 
 which  parameters are  a centroid line energy $E_{line}$, the width of the line $\sigma_{line}$  
and the normalization $N_{line}$ to fit the data in the 6 -- 8 keV  energy range.  
%(see Fig. \ref{BeppoSAX_spectra}). 
 We also use   the interstellar absorption with a column density $N_H$.
%  in the model.
It should be noted  that we  fixed certain parameters of the $COMPTB$ component: 
$\gamma=3$ (low energy index of the seed photon spectrum) and $\delta=0$ because we neglect an efficiency  of the   bulk inflow effect vs the  thermal Comptonization   for  NS  GX~3+1. 
We apply a value of hydrogen column $N_H=1.6\times 10^{22}$ cm$^{-2}$, which was found by~\citet{ooster01}.  
%{\it }
%%%% END

Initially, we have tried  a model consisting of an absorbed thermal component ($bbody$) and a thermal Comptonization 
component ($COMPTB$)  but this model gave a poor description of data. Significant 
positive residuals around $\sim$~6.5 keV suggest
%and negative residuals near $\sim$~7 keV (and $\sim$~10 keV ????) 
the presence of fluorescent iron emission line. %and corresponding absorption edge. 
The addition of $Gaussian$ line component at 6.4 keV %and absorption edge component ($edge$) at 7.1 keV 
considerably improves fit quality and provides a statistically acceptable $\chi^2_{red}$.

At the first time the fluorescent iron emission line in GX~3+1  was detected by \cite{ooster01} using  $Beppo$SAX 
on 1999 August, 30 (id=20835001). 
%by Oosterbroek et al. (2000)
% in spectrum of GX~3+1 observed 
\cite{ooster01} successfully described this emission feature with 
the $Gaussian$ line model and  they used a model consisting of thermal component 
(dominating energy release around 1 keV) and a thermal Comptonization  {\it tail} to describe the 0.1 -- 50 keV continuum.  However, 
they needed to add a 2\% systematical uncertainty to LECS and MECS data to obtain acceptable 
$\chi^2_{red}$. We  investigate a possibility to further improve  a  quality of the fit.

%%We found   that the emission line feature  is  quite broad and it is much wider than 
%%the instrumental response whose width is smaller than 0.02 keV
%%\footnote{See ftp://heasarc.gsfc.nasa.gov/sax/cal/responses/98\_11}.
%%Thus  we   include a simple {\it Gaussian} component, 
%% whose  parameters are  a centroid line energy $E_{line}$, the width of the line $\sigma_{line}$  
%%and the normalization $N_{line}$,  in the model to fit the data in the 6 -- 8 keV  range.  
%%%(see Fig. \ref{BeppoSAX_spectra}). 
%%We also include  the interstellar absorption with a column density $N_H$ in the model.
%%It should be noted  that we  fixed certain parameters of the $COMPTB$ component: 
%%$\gamma=3$ (low energy index of the seed photon spectrum) and $\delta=0$ because we neglect an efficiency  of the  
%%bulk inflow effect vs the  thermal Comptonization   for  NS  GX~3+1.

\begin{figure}[ptbptbptb]
\includegraphics[scale=0.98, angle=0]{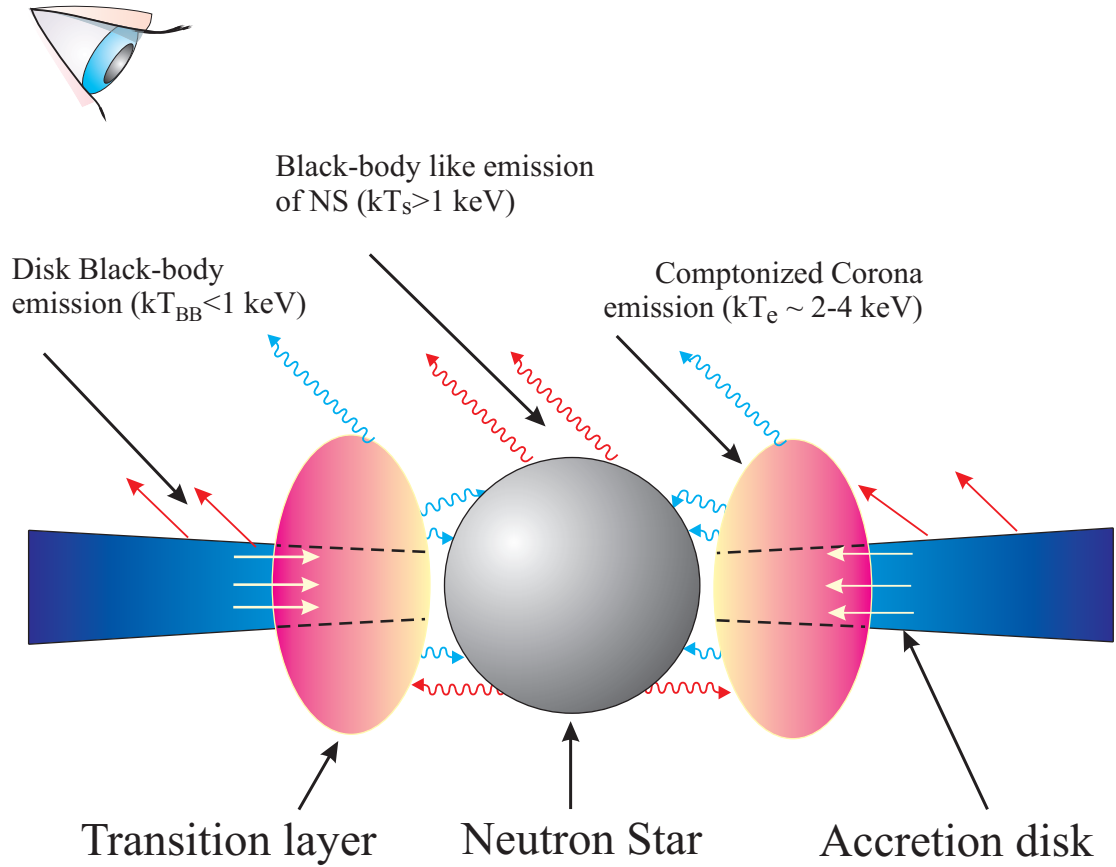}
\caption{A suggested  geometry of GX 3+1.   Disk and neutron star soft photons are 
up-scattered off   hotter plasma of the transition layer located 
between the accretion disk and NS surface.  Some fraction of these 
photons is seen directly by the Earth observer. Red and blue photon trajectories correspond to soft 
and hard (upscattered) photons respectively. 
%Upper panel presumably describes  the spectral formation 
%in the $faint$ phase %low state 
%when the TL temperature is higher than that in the bright phase. %high state. % (see bottom panel). 
}
\label{geometry}
\end{figure}

In Figure~\ref{geometry}  we illustrate  our spectral model 
as a basic model for  fitting  the {\it Beppo}SAX and {\it RXTE}  spectral data 
for GX~3+1.
We assume that accretion onto a neutron star takes place when the material passing through 
the two main regions:  a geometrically thin accretion disk [the standard Shakura-Sunyaev 
disk, see \citet{ss73}]
and the  transition layer (TL), where NS and  disk soft photons   are  upscattered off hot electrons. 
In other words, in our picture, the emergent thermal Comptonization spectrum is  formed in the  TL, 
where thermal disk  seed photons and soft photons from the neutron star  are upscattered off the relatively hot plasma (electrons). 
Some fraction of these seed soft photons can be also seen directly.
% by the Earth observer. 
Red and blue photon 
trajectories shown in Fig. \ref{geometry} correspond to soft 
(seed) and hard (up-scattered) photons respectively. 

\begin{figure}[ptbptbptb]
\includegraphics[scale=0.9,angle=0]{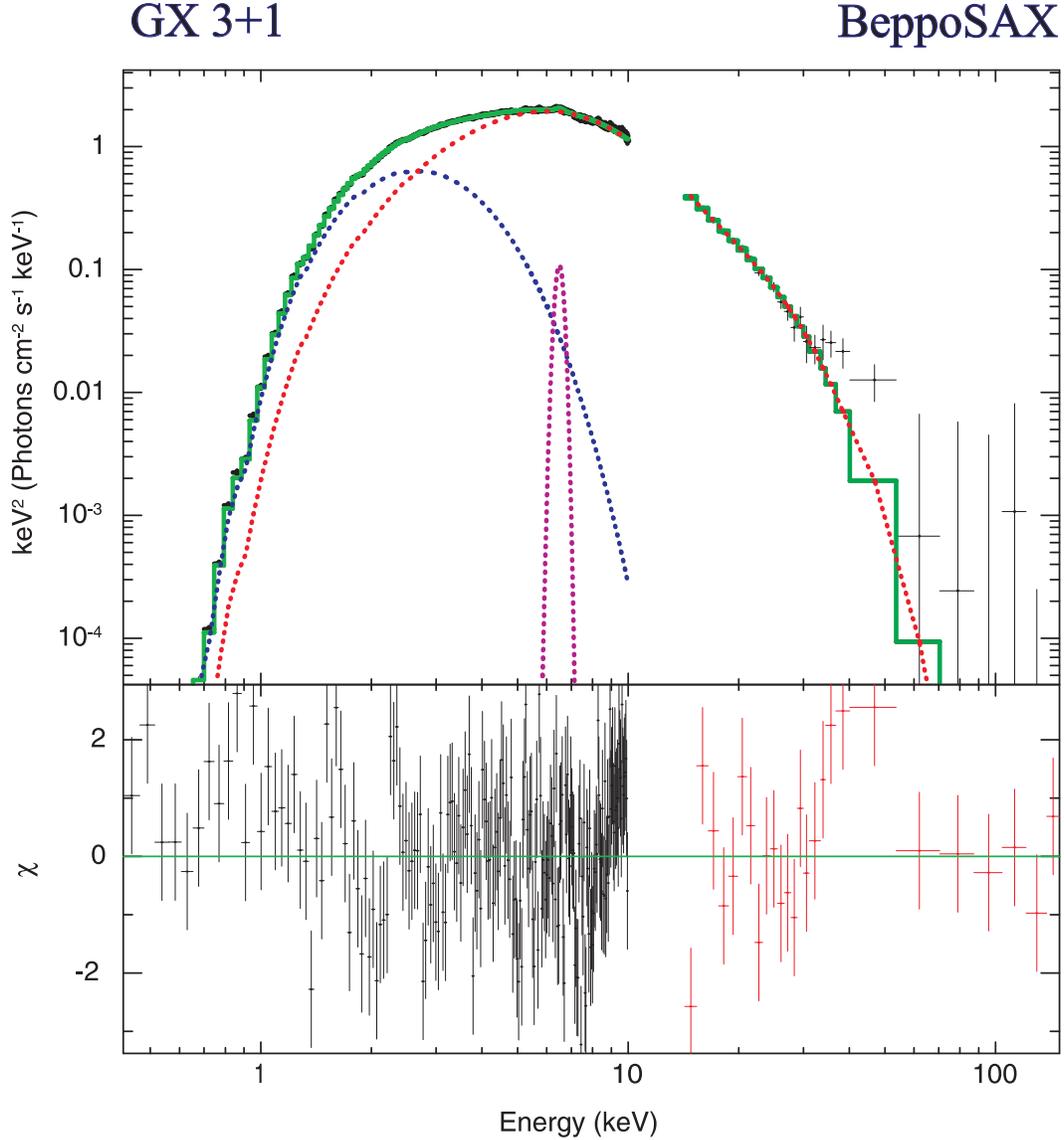}
\caption{$Top:$ the best-fit spectrum of GX~3+1 %during {\it Intermediate state} events 
 in $E*F(E)$ units using {\it Beppo}SAX observation  20603001 carried out on 28 February -- 1 March 1999.   
The data  are presented by crosses and the best-fit spectral  model   {\it wabs*(blackbody+COMPTB+Gaussian)} 
by green line. 
%{\bf Lenochka, why here is a green line only shown  to photon energies below 10 keV only?}
 The model components  are shown by blue, red and crimson lines for {\it blackbdody}, 
{\it COMPTB}  and {\it Gaussian} components respectively. 
%$Top:$  the best-fit spectrum of GX~3+1 %during {\it Intermediate state} events 
% in $E*F(E)$ units using {\it Beppo}SAX observation  20603001 carried out on 28 February -- 1 March 1999.   The data are presented by crosses and the best-fit spectral  model   {\it wabs*(blackbody+COMPTB+Gaussian)} by green line. The model components  are shown by  red, crimson and  blue lines for {\it blackbdody}, {\it COMPTB}  and {\it Gaussian} components respectively. 
 $Bottom$: $\Delta \chi$ vs photon energy in keV. 
The best-fit model parameters are 
$\Gamma$=1.99$\pm$0.07, $T_e$=3.68$\pm$0.05 keV and $E_{line}$=7.4$\pm$0.1 keV (reduced $\chi^2$=1.08 for 457 d.o.f) 
(see more details in Table 3). 
}
\label{BeppoSAX_spectra}
\end{figure}

\begin{figure}[ptbptbptb]
\includegraphics[scale=0.95,angle=0]{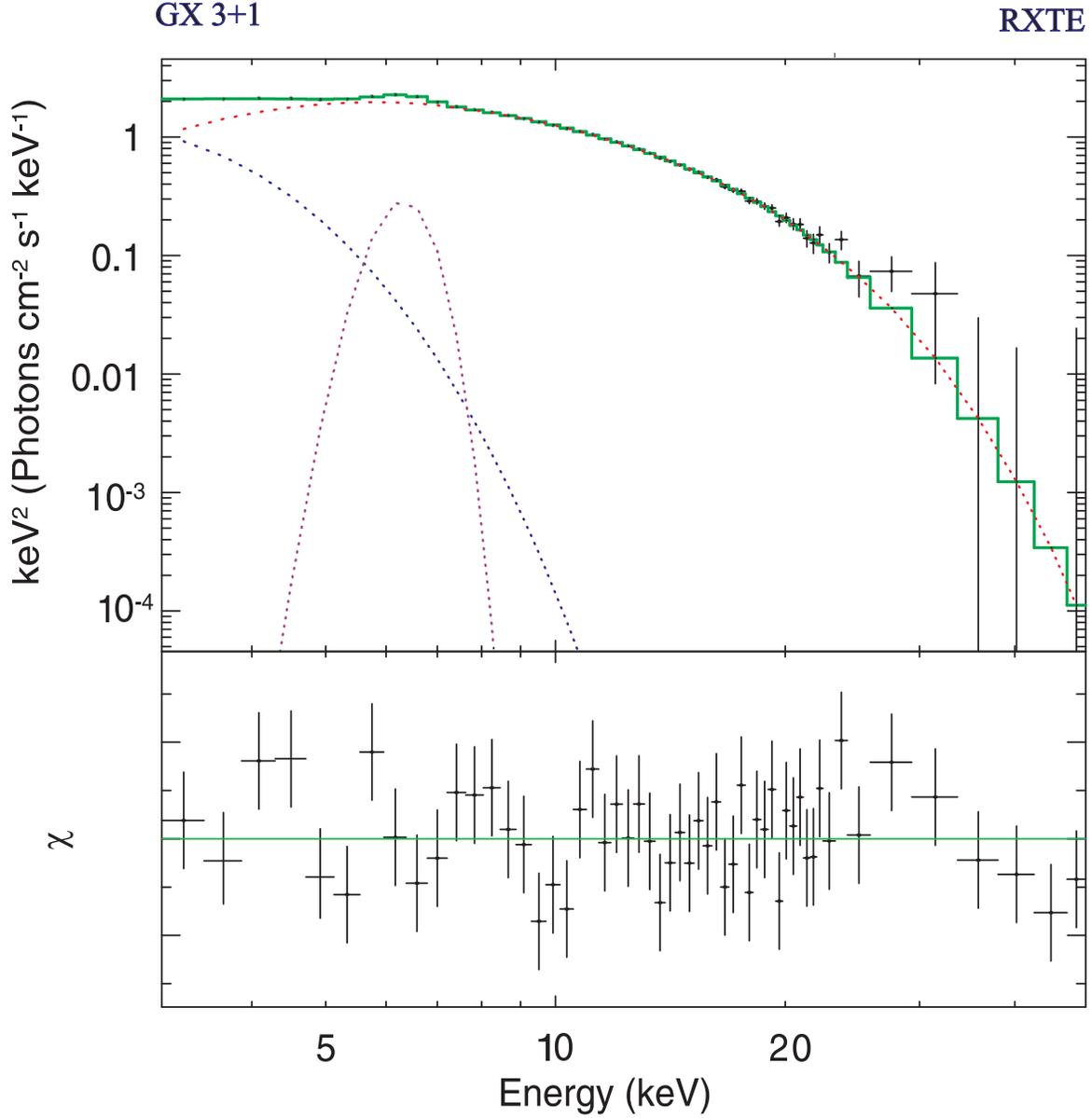}
\caption{ {\it Upper} panel: A typical $E*F(E)$ spectral diagram  of GX~3+1
for PCA/{\it RXTE} observation 94307-05-01-000 on 2010, 1 September (MJD 55431.62)  during the 
$faint$ phase (green line). %$low$ state  .
The spectral model, presented by its  components is shown
 by dashed red, blue, and purple lines for $COMPTB$, $blackbody$ and $Gaussian$ components respectively.
{\it Bottom} panel: $\Delta\chi$ vs  photon energy in keV. The best-fit model parameters
are $\Gamma=2.04\pm 0.04$, $kT_e=3.16\pm0.05$ keV and $E_{line}=6.59\pm 0.05$ keV
(reduced $\chi^2=0.87$ for 67 d.o.f.) (see more details in Table 4). 
}
\label{rxte_hard_state_spectrum}
\end{figure}

We show examples of X-ray spectra in  Figs.~\ref{BeppoSAX_spectra} -- \ref{rxte_hard_state_spectrum}  for $Beppo$SAX  and {\it RXTE}  data respectively.
%and in Fig. 
%-$\ref{rxte_soft_state_spectrum}  
%(for   ). 
Spectral analysis of {\it Beppo}SAX and  {\it RXTE}  observations indicates  that X-ray  spectra of GX~3+1 can be   
described by a model with a  Comptonization component  represented by  the $COMPTB$ model. %(Farinelli et al, 2008)
Moreover,  for  broad-band {\it Beppo}SAX  observations this spectral model  component is  modified  by  
photoelectric absorption at low energies. 

On the $top$ of Figure~\ref{BeppoSAX_spectra} we demonstrate  the best-fit {\it Beppo}SAX
spectrum of GX~3+1 in units of 
$E*F(E)$ ($top$) [where $F(E)$ is energy flux in erg, per keV and per second]  using  our   model for  {\it Beppo}SAX observation (id=20603001) carried out on 28 February -- 1 
March 1999.    The data are presented by crosses and the best-fit spectral  model   {\it wabs*(blackbody + COMPTB + Gaussian)} 
by green line. The model components  are shown by blue, red and crimson lines for {\it blackbdody}, 
{\it COMPTB}  and {\it Gaussian} components respectively. 
On the $bottom$ we show $\Delta \chi$ vs    photon energy in keV. 
The best-fit model parameters are 
$\Gamma$=1.99$\pm$0.07, $kT_e$=3.68$\pm$0.05 keV and $E_{line}$=7.4$\pm$0.1 keV (reduced $\chi^2$=1.08 for 457 d.o.f) 
(see more details in Table 3).
In particular, we  find that  an addition of  the soft thermal   
component with   temperature $kT_{BB}=$0.5$-$0.7 keV to the model  significantly improves  the fit quality of the  
{\it Beppo}SAX  spectra. For the {\it Beppo}SAX data 
%observations 
(see Tables 1, 3) we find that the spectral index $\alpha$ 
is %around 1 
of 1.03$\pm$0.04
(or the corresponding photon index $\Gamma=\alpha+1$ is %about 2
 2.03$\pm$0.04). 

Unfortunately  {\it RXTE} detectors do not provide well calibrated spectra  below 3 keV  while the  
broad energy band of {\it Beppo}SAX telescopes allows us to determine  the parameters 
of {\it blackbody} components  at low energies.  
Thus, in order to fit the {\it RXTE} data  we have to fix the  temperature of {\it blackbody} component at a value of 
$kT_{BB}=$0.6 keV obtained as an upper limit  in  our   analyze of   the {\it Beppo}SAX data.
% from  4U~1728-34. 
The  best-fit spectral parameters   using {\it RXTE}  observations are presented in Table 4. 
In particular,  we find that electron temperature $kT_e$ of the  $COMPTB$ component varies from 2.3 to 4.5 keV,  
while the photon   index $\Gamma$  is almost constant ($\Gamma=1.99\pm 0.02$)
 for all observations.  It is worth noting that the width $\sigma_{line}$ of $Gaussian$ component does not  vary significantly   and it is  in the range of 0.5 -- 0.8 keV.

Color temperature $kT_{s}$  of $COMPTB$ component changes from  1.2 keV to  1.7 keV,
 which is consistent with that using  the {\it Beppo}SAX data set 
of our analysis (see Table 3) and previous studies by  \citet{ooster01}, \citet{denhart03} and  \citet{ch06}. 
We should also emphasize  that the  temperature of the seed photons $kT_s$  of the $COMPTB$ component usually  increases up to 1.7 keV in the
fainter phases %low luminosity state 
and generally decreases to 1.2 keV in the $bright$ phases. %high luminosity state. 

%We use a value of hydrogen column $N_H=1.6\times 10^{22}$ cm$^{-2}$, which was found by~\citet{ooster01}.  
%%%Systematic error of 0.5\% has been applied to all analyzed {\it RXTE}  spectra.
In  Figure~\ref{rxte_hard_state_spectrum} we show an example of 
the best-fit {\it RXTE} spectrum of GX~3+1 for the  fainter luminosity phases %low 
 %state 
and the  residuals 
% the spectrum in counts  units 
({\it bottom panel}) with 
$\Delta\chi$ for the 94307-05-01-000 observation. 
Blue, red and  purple lines stand for 
$blackbody$, $COMPTB$ and $Gaussian$ components, respectively.

\begin{figure}[ptbptbptb]
\includegraphics[scale=0.9,angle=0]{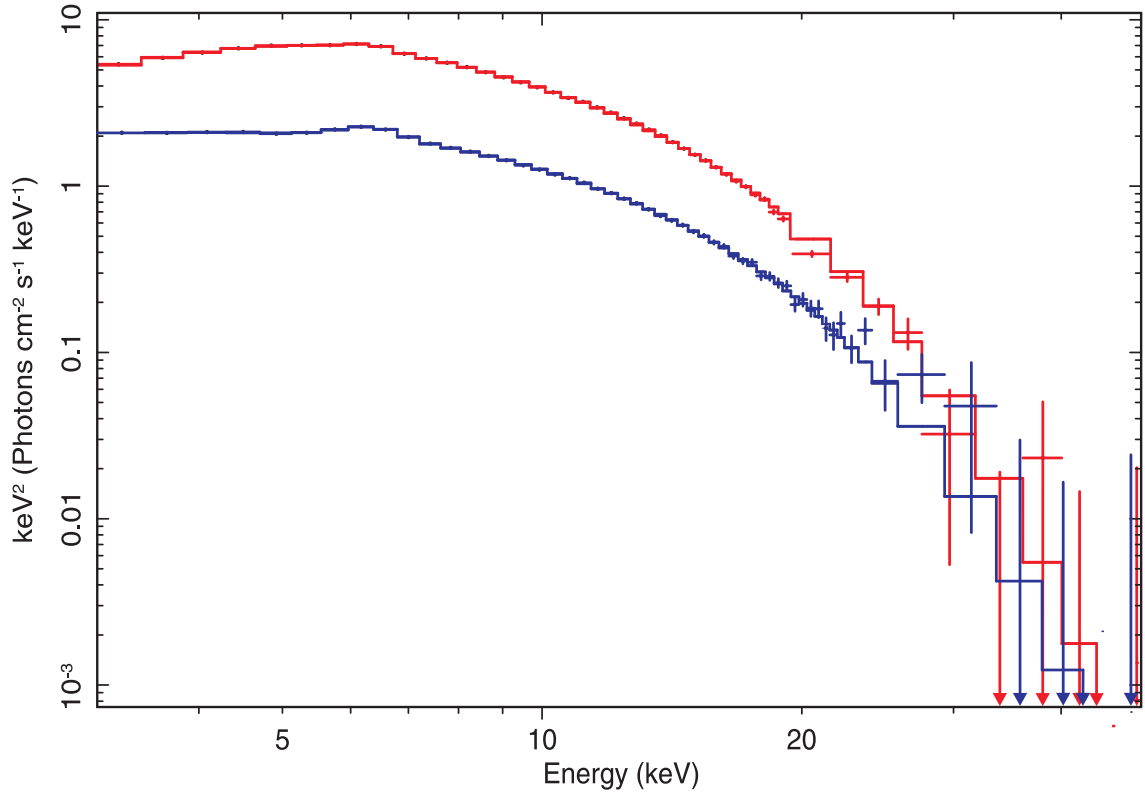}
\caption{Examples of X-ray spectra, presented as $E*F(E)$ spectral diagrams,  of GX~3+1 during 
$faint$ phase %$low$ state 
(94307-05-01-00, $blue$) and 
$bright$ phase %$high$ state 
(60022-01-13-01, $red$) detected with {\it RXTE} on MJD 55440.62 and 52544.48 
respectively.
}
\label{rxte_soft_state_spectrum}
\end{figure}

In  Figure~\ref{rxte_soft_state_spectrum} we also show examples of typical photon spectra $E*F(E)$ spectral diagrams of GX~3+1 during 
the $fainter$ phase %low$ state 
(94307-05-01-00, $blue$) and 
the $brighter$ % high$ state 
phases (60022-01-13-01, $red$) detected with {\it RXTE} on MJD 55440.62 and 52544.48 
respectively.
The adopted spectral model shows a very good fidelity throughout
all data sets used in our analysis. Namely, a value of reduced
$\chi^2_{red}=\chi^2/N_{dof}$, where $N_{dof}$ is a number of degree of freedom, 
is  less or about 1.0 for most observations. For a small 
fraction (less than 2\%) of spectra with high counting statistics
$\chi^2_{red}$ reaches 1.5. However, it never exceeds our rejection 
limit of 1.7. 
%%%%BEGIN
%{\it
Note that the energy range for the cases, in which we obtain  the  poor fit statistic 
(two among 101 spectra with $\chi^2$=1.7 for 67 dof),   are related to the iron line region.  
%Possibly, it is caused with complexity of iron line shape and relatively poor 
%energy resolution of RXTE. 
It is possible that the shape of iron line is more complex than a simple Gaussian (i.e. a  blend of different 
energies, presence of the edge, or broadening by Comptonization). The fits tend to favor a broad line  
(see Table 4), which might be caused by Comptonization. However, this possible complexity is not well  
constrained by our data. 

It is worth noting that we find some differences 
between our values of the best-fit model parameters and those in the literature.
%  for the same set of the  $Beppo$SAX observations. 
  In particular, the photon index  $\Gamma$, estimated by \citet{ooster01} 
  %Di Salvo et al. (2000)
   for observation  id=20835001, is 1.60$\pm$0.25. 
%while our value of $\Gamma=1.9\pm0.2$. 
This  discrepancy in index values  can be a result of   using   slightly  different  spectral models by us and \citet{ooster01}. 

Thus   using the broad band 
{\it Beppo}SAX observations we can accurately  determine all of the  parameters 
of our spectral model 
%the model components of GX~3+1 spectrum. 
while  using  the extensive observations of GX~3+1 by {\it RXTE}  we  are able to  
investigate the overall pattern of the source behavior during the spectral transitions 
in the 3 -- 50 keV energy range.  

\section{Overall pattern of X-ray properties 
%versus transient behavior 
\label{evolution_1}}
%\section{Evolution of X-ray  properties during state transitions \label{evolution}}

\subsection{Evolution of X-ray  spectral properties during transitions \label{evolution}}

As was mentioned above, at time scales larger that one minute, GX~3+1 exhibits two kinds of variability,  $slow$ and $mild$. 
The former one  ($slow$) has a time scale of order years. This $slow$ variability are seen 
in the $faint$ phases %on luminosity 
%so called {\it low} state (LS) 
and {\it bright} phases %high} state (HS) 
which are related to low and high luminosities, respectively. On the other hand the $mild$
variability has a time scale of order of days and modulation depth in the 3 -- 10 keV band 
is typically 20\%. The ASM (2 -- 12 keV) mean flux correlates with COMPTB normalization ($N_{COM}$) 
and tends to anticorrelate with the electron plasma temperature of Compton cloud (CC) $T_e$ (see Fig.~\ref{lc_1998}). 
Such $mild$ variability is detected for both the $fainter$ and $brighter$ phases  %$low$ and $high$ states 
for GX~3+1. It should be noted that 
the COMPTB normalization changes are  larger in the  {\it bright} phase %high} state 
than that  during the  {\it faint} phase, %low} state, 
while the 
electron temperature $T_e$ variations are almost the same for both phases.
%states.

One can relate $slow$ and $mild$ variabilities of GX~3+1 to slow and mild changes of mass 
accretion rate, respectively. The $slow$ variability by factor 4 %2 
has been seen in the 1996 -- 2010 observations by 
ASM/{\it RXTE}.  The same kind of changes of the flux was also  observed  in the  earlier observations by  \cite{mak83}.  
%(Makishima et al., 1983).
 In turn, in the next section it will be shown that the $slow$ variability can be related to  
%local 
transitions between the {\it brighter} and {\it fainter} phases
%{\it lower banana} and {\it upper banana} states
 along with small variations of the electron temperature $kT_e$. 
 %the electron temperature ($kT_e$) changes.

We found that  the X-ray spectra of GX 3+1 over the $bright$ and $faint$ phases %$low$ and $high$ states 
are quite stable. Moreover, we 
detected a constant 3 -- 50 keV spectral shape over the $slow$ and $mild$ variability stages. The best-fit parameters 
are listed in Table 4. The {\it faint/bright phase} %LS-HS 
transitions  are characterized  
by the  spectra with an almost constant spectral  index $\alpha$ of 1, or photon index $\Gamma$ of 2  (see Fig.~\ref{evolution_lc_all}). 
We have also  established  common characteristics of the  rise-decay spectral transition  of GX~3+1 
based on their  spectral parameter evolution of X-ray emission  in  the energy range from 3 to 50 keV  
using PCA/{\it RXTE} data.  
In Figures \ref{rxte_hard_state_spectrum}$-$\ref{rxte_soft_state_spectrum} we present typical examples  
of the {\it RXTE} bright and faint phase %low and high state  
spectra for GX~3+1. In fact, one can clearly see  from these 
Figures % \ref{rxte_soft_state_spectrum}$-$\ref{rxte_hard_state_spectrum}  
that the normalization of the thermal  component is a factor of  2 higher in the $bright$ phase %high state  
than that in the faint phase, %low state, 
although  the photon indices $\Gamma$ for each of these spectra
% only vary  from 1.9 to 2.1 and  
are  concentrated  around 
%$\Gamma$=
2 (see 
 %that distribution of  
%$\Gamma$ on the {\it  left-hand} panel of
 Figs.~\ref{lc_1998} -- \ref{evolution_lc_all}). 

In fact, we test the hypothesis  of  $\Gamma_{appr} \approx 2 $ using  $\chi^2$-statistic criterion.
% by minimization of corresponding 
%residuals for all measurements. 
We calculate the distribution of $\chi^2_{red}(\Gamma_{appr})=\frac{1}{N}\sum_{i=1}^N\left(
\frac{\Gamma_i-\Gamma_{appr}}{\Delta\Gamma_i}
\right)^2$  versus of $\Gamma_{appr}$
% on the {\it right-hand} panel of  Figure~\ref{hist}.
and we  find  a sharp minimum of  function $\chi^2_{red}(\Gamma_{appr})$ around 1 
which takes place in the range of 
%$\Gamma_{appr}=1.99\pm0.02$ for d.o.f=126 with null hypothesis probability $10^{-8}$.
%It is worth noting that 
$\Gamma_{appr}=1.99\pm0.01$ with a confidence level of 67\%  and 
$\Gamma_{appr}=1.99\pm0.02$ with a confidence level 99\% for 101 %99 
d.o.f. (see the similar Figure of  $\chi^2_{red}(\Gamma_{appr})$ for 4U 1728-34 in ST11).
%It is important to emphasize that the photon index  $\Gamma$  is also independent of the normalization of $COMPTB$,  $L_{39}^{soft}/d^2_{10}$ and the plasma temperature of Compton cloud $T_e$
% when both of these parameters change by a factor 5 at least 
%(see Fig.
% \ref{outburst_index_norm}$-
% \ref{outburst_index_temperature1}).
 Using {\it Beppo}SAX  data  FT11  suggested  that the  photon index $\Gamma$ is approximately 2 for many   
NS binaries which are observed in  different spectral states.  FT11 characterize  the spectral state 
by a value of electron temperature $T_e$  and they show that $\Gamma=2\pm 0.2$ (or 
$\alpha=1\pm 0.2$)  when  $kT_e$ changes from 2.5 to 25 keV. 

A number of X-ray flaring episodes of GX~3+1 has been  detected with {\it RXTE} during  2001$-$2002
($R4$ set) with a good rise-decay coverage.  We have searched for  common 
spectral and timing features which can be revealed during these spectral transition episodes.  
We present the combined results of  the spectral analysis of these observations  using 
 our  spectral model  $wabs*(blackbody+COMPTB+Gaussian)$  in
  Figures \ref{lc_1998}$-$\ref{evolution_lc_all}.   
 ASM/{\it RXTE}  count rate is shown on the top panel of these Figures. 
Further, from the  top to the bottom,  we show the  model flux in two energy bands  3 -- 10 keV 
({\it blue points})   
and 10 -- 50  keV  ({\it crimson  points}).  In the next panel  we show 
a change of the TL  electron temperature $kT_e$. One can clearly see the 
{\it low amplitude} spectral transition {\it on time scales of $\sim$ 1 -- 2 days} from the $brighter$ phase %high state 
to the $faint$ phase %low state   
during the time period from MJD 52000 to  
MJD 52200 while $kT_e$ only varies from 2.3 keV to 4.5 keV during this transition.  

Normalizations of  the $COMPTB$ and   $blackbody$ component ({\it crimson} and  {\it blue}  points respectively) 
 are shown in the next panel  of 
Figs.~\ref{lc_1998} and \ref{evolution_lc_all}. 
In particular, one can see from Figures \ref{lc_1998}$-$\ref{evolution_lc_all}  
how the $COMPTB$ normalization $N_{COM}$ correlates with  
the variations of  ASM count rate and  the model flux in 3-10 keV energy band. 
On the other hand, the normalization of the $blackbody$ component $N_{BB}$ is almost constant 
except at   the mild variability  episode peak, when $N_{BB}$ increases from 0.02 to 0.14 
(see blue points   in Figure \ref{lc_1998}    at MJD=52130 and 52170).
Moreover  these  spectral variability  transitions are related to 
a noticeable  increase of  flux in the 3 -- 10 keV energy range  and corresponding decrease of  flux 
that takes place  in the 10 -- 50 keV energy range (see the second panels from above in 
Figs.~\ref{lc_1998} -- \ref{evolution_lc_all}).

%The spectral  index $\alpha$ ($\alpha=\Gamma-1$) is presented in the bottom panels of 
% Figures \ref{lc_1998}$-$\ref{evolution_lc_all}. 
 %One can see that  
  
The illumination  fraction $f$  varies from 0.1 to 0.9 as seen
% from  on the right hand bottom  panel of  
from  Figure \ref{T_e_vs_f_comp} %{outburst_index_temperature1} 
while the index $\alpha$  only slightly varies with time around 1 (or $\Gamma\sim 2$) (see Figs. \ref{lc_1998}$-$\ref{evolution_lc_all}).
%Consequently, 
%Comptonization  fraction $f$ shown in the right hand  panel of Fig.~\ref{outburst_index_norm} is 
 %high in most of cases. 
 However in most  cases the soft disk  radiation  of GX~3+1 
 is subjected to reprocessing in a Compton cloud  and only some fraction of 
the soft photon emission component ($1-f$) is directly seen by the Earth observer. 
Note that $f$ changes  with COMPTB normalization (see Fig.~\ref{T_e_vs_f_comp}, %{outburst_index_temperature1}, 
the inner panel in the left-hand upper corner).
The energy spectrum of GX~3+1 during almost all states is dominated by a Comptonized component %seen  as a  power-law hard emission in the energy range from 3 to 60 keV, 
while  the direct soft photon emission is %not seen 
always weaker and detectable in the flaring episodes  only  
(see also Figs.~\ref{lc_1998}$-$\ref{evolution_lc_all}).  

Note that for BHs a definition of spectral transition  involves a change of photon index 
$\Gamma$ (see e.g. ST09).  However there is no   one-to-one correspondence between $\Gamma$  and 
cutoff  (or efold) energy $E_{fold}$. \citet{ts10} demonstrate using {\it RXTE} data for BH binary
 XTE J1550-564 that   $E_{fold}$ decreases when $\Gamma$ increases from 1.4 to $2.1-2.2$ until    $\Gamma$ reaches 2.2   and then   $E_{fold}$ increases.  
%There spectral hardness during burst spectral transition. 
Thus {\it for a BH the main parameter used for the spectral transition definition    
is a {\it variable} photon index $\Gamma$ which monotonically increases when the source goes into the $bright$ phase %high state
}.

\begin{figure}[ptbptbptb]
\includegraphics[scale=0.98,angle=0]{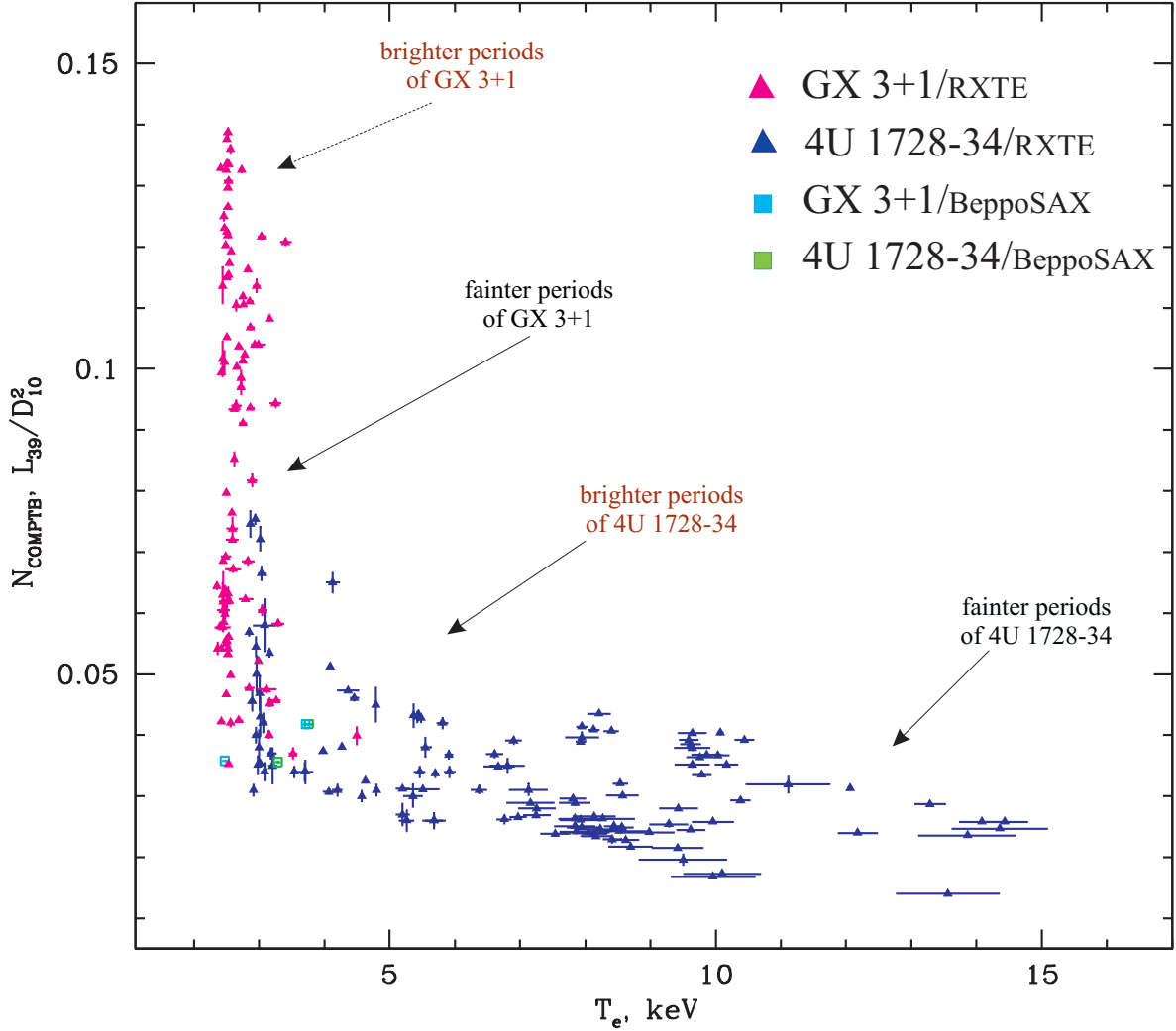}
\caption{COMPTB normalization measured in $L^{soft}_{39}/D^2_{10}$ units 
versus electron temperature $T_e$ (in keV) obtained using the best-fit  spectral model $wabs*(blackbody+COMPTB+Gaussian)$ for {\it atoll} sources GX~3+1 ($pink$) and 
4U~1728-34 ($blue$, taken from ST11)  for  {\it RXTE} data and    
%whereas 
 {\it bright blue} and $green$ points for $Beppo$SAX data. Mass accretion rate continuously increases along this correlation  from the right to the left. 
 % in terms of the best-fit
 % analysis 
%in frame of %the same for both sources 
%. 
%The areas of ``brighter/fainter periods'' of GX~3+1 
%and 4U~1728-34 are indicated for demonstration continuous increasing of mass accretion rate for both 
%objects along the ``common'' track pass, formed by two individual tracks.
} 
%{\it From Top to Bottom:} ASM light curves for ``atoll'' sources GX~3+1 ($pink$); for GX~3+1 ($blue$); 
%diagrams $\Gamma$ vs COMPTB-Normalization ($left$), vs Comptonized fraction $f$ ($middle$) and vs the electron temperature $T_e$ for GX~3+1 and 4U~1728-34 respectively.
%diagrams $\Gamma$ vs COMPTB-Normalization ($left$), vs Comptonized fraction $f$ ($middle$) and vs the electron temperature $T_e$ for 4U~1728-34.
%In two $bottom$ panels $pink$ and $blue$ points correspond to {\it RXTE} data whereas 
%$green$ and $red$ points correspond to $Beppo$SAX data for GX 3+1 and 4U 1728-34 respectively.}
%\label{outburst_index_temperature2}
\label{norm_T_e}
\end{figure}

It is important to emphasize once again that in the NS binary GX~3+1  the transition  from  the $faint$ phase 
%low state 
to the $bright$ phase % high state  
takes place  
when COMPTB normalization  $N_{COM}=L_{39}^{soft}/D^2_{10}$ %electron (plasma)  temperature 
changes  from  0.04 to 0.15.   
%$-$\ref{evolution_lc_3}).       
%6, 7). 
Thus, 
%following FT11 suggestion, 
we define   %spectral 
the NS spectral  transition   
%can be considered (defined, suitable to consider) 
in terms of the COMPTB normalization. %electron temperature $T_e$ of the Compton cloud (TL).
 In this case 
the  faint phase %low state 
is characterized by the low  normalization value  while 
the $bright$ phase %high state 
is related to the high  normalization value. 
% Note that the COMPTB normalization %electron temperature $T_e$ 
%is a directly measurable quantity. 
In Figure~\ref{norm_T_e} %{outburst_index_temperature2} 
we demonstrate the dependence of COMPTB normalization  $L_{39}^{soft}/D^2_{10}$ on 
%the electron temperature
 $kT_e$ using 
these best-fit parameters for GX 3+1 and 4U 1728-34 for the {\it Beppo}SAX and {\it RXTE} data.  From this 
Figure %\ref{outburst_index_temperature2} 
one can clear see  a monotonic behavior   $N_{COM}$ vs $kT_e$, namely  $kT_e$ decreases when the soft  flux 
increases. It is worth noting that the $kT_e$ values obtained using GX 3+1 data for {\it Beppo}SAX and {\it RXTE} reache the asymptotic value of about 2.5 keV at high values of the soft flux ($N_{COM}>0.05$).
%%% BEGIN

\begin{figure}[ptbptbptb]
\includegraphics[scale=.9,angle=0]{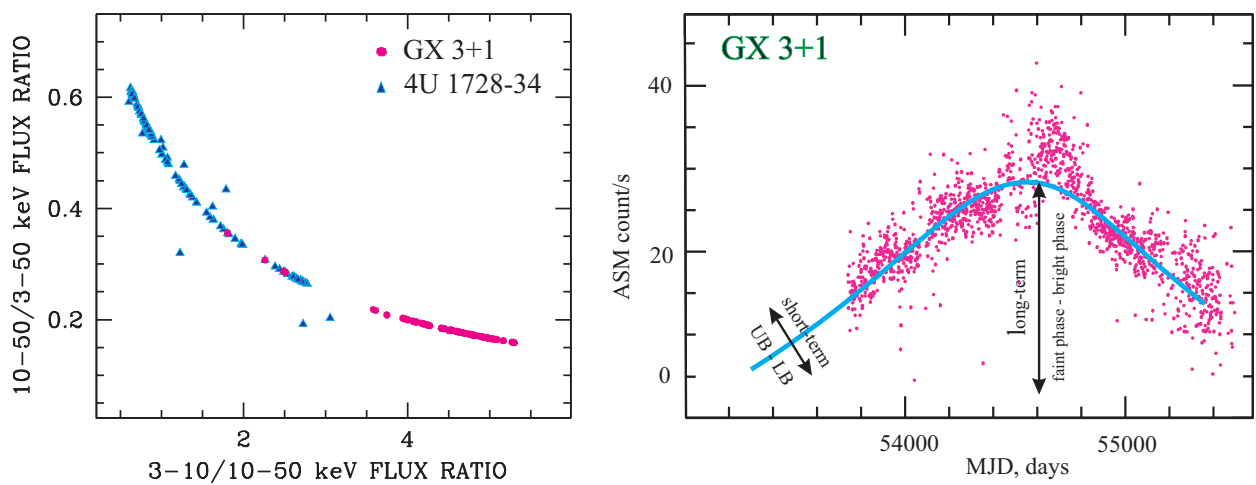}
\caption{{\it Left:} The  {\it color-color diagram} %[$\frac{3-10 keV}{10-50 keV}$ %
[3 -- 10/10 -- 50 keV flux ratio versus 10 -- 50/3 -- 50 keV flux ratio] of GX~3+1 ($pink$) and 
4U~1728-34 ($blue$) during faint-bright %LS -- HS 
transitions (long-term variability). 
{\it Right:} Fragment of ASM light curve of GX~3+1 %is shown 
which shows 
%for (schematic) demonstration (illustration) of 
two types  of flux variability. One is a long-term  trend (from faint to bright) %LS -- HS) %with high amplitude
related to COMPTB normalization changes and another one  
 is  short-term time variations (UB -- LB) %with low amplitude 
related to   the electron temperature  changes.
%of Compton cloud changing
% are indicated. 
The blue line presents
a mean count rate  and indicates to  a long-term variability of GX~3+1.
% soft flux.
%$power$ diagram of GX~3+1  observed on September 1, 2010 
%(94307-05-01-00 {\it RXTE} observation, MJD=55441).
}
\label{color_diagram}
\end{figure}

To demonstrate transition properties of GX~3+1 in terms of flux (or luminosity) we define the
hard color (HC) as a ratio of the flux in the 10 -- 50 keV to that in the 3 -- 50 keV energy band, 
and the soft color (SC) as a ratio of the flux in the 3 -- 10 keV to that in the 10 -- 50 keV energy range. 
Plotting HC versus SC we created $our$ $color-color$ diagram (CD, see the left panel of Fig.~\ref{color_diagram}
%10
) for two atoll 
sources GX~3+1 ($pink$) and 4U~1728-34 ($blue$). As it appears from this Figure the tracks of these two 
sources display  a smooth and continuous (monotonic) function, pointing  out the similar
physical mechanism of hard/soft flux transition during a long-term source evolution for these  two objects.
In Fig.~\ref{color_diagram}
%10  
($right$ panel) we demonstrate a  fragment of ASM light curve of GX~3+1 which illustrates  two  types of flux variability. 
The long-term time trend (from bright  to  faint) %(LS -- HS) %with high amplitude associated 
is related to  COMPTB normalization changes and while the second one shows   %direction 
 short-term time variations (UB-LB)  %with low amplitude 
related to  the Compton cloud  electron temperature variations.  The blue line displays a mean count rate and indicates to  long-term variability of GX~3+1 flux.
%the power spectra in power$\times$frequency versus frequency ($\nu\times P(\nu)$) 
%representation.GX~3+1 show a ``$\Gamma$''- (or inverted ``J''-like) shape in our CD. Here the track of GX~3+1 
%contains two branches: slightly inclined (horizontal) part (branch) at the central (left) top part of our CD and 
%vertical branch at the left part. Thus, the track of GX~3+1 represents two branched curve with smooth 
%and continuous transition from the inclined to vertical part. This track is associated with $banana$ 
%states in accordance with classification by~\cite{hasinger89}. Specifically, the inclined (horizontal)  
%branch is associated with Lower banana (LB) state and vertical branch is correspond to Upper banana (UB) state. 
%This state identification is supported by combination of spectral (see Sect.~\ref{evolution}) and timing properties 
%(Sect.~\ref{transitions}) in agreement with $atoll$--$Z$ scheme. 
Note that the track of Fig. ~\ref{color_diagram} ($left $ panel)
%in whole,
reflects a long$-$term evolution of GX~3+1.
 %whereas $slow$ evolution thereof is presented as reciprocate along the track without visual 
%effect (not perpendicular to a track pass).

It is worth  noting that among  all NSs 
%only  show spectral transitions and  
only a few of 
$atoll$ and Z-sources  
demonstrate  a full 
%(complete,  
track on CD in a wide range of luminosity.
%%% END
%Not all NSs show flares.  Only  a few NS binaries (such as Z and $atoll-$sources)  display   spectral  transitions during the bursts. 
For example, $atoll$ sources, such as 4U~1728-34, usually show a wide range of spectral states during transitions which are related 
to changes of the  total luminosity and mass accretion rate. 
One can   establish   a substantial 
difference between a NS and a BH due to these flare episodes when a  source evolves from the 
$faint$ phase %low state 
to the $bright$ phase %high state 
and when the  plasma temperature of a Comptonized region  changes remarkably
(like in 4U 1728-34 from 2.5 keV to 15 keV during IS -- B states). On the other hand  %In contrast 
%to 4U~1728-34,   
GX~3+1 
shows significant changes in the total luminosity but with only a slight   variation of electron  temperature $kT_e$ 
in its $banana$ state. However,  the  photon index $\Gamma$ stays around  a value of 2  and is independent of the soft photon luminosity 
both in  the faint phase %low state 
and  the bright phase. %high state  
%and $T_e$ (see Fig. \ref{outburst_index_temperature1}) similar to that in  4U~1728-34.

\subsection{Timing properties during LB-UB  transitions \label{transitions}}

The {\it RXTE} light curves have been analyzed using the {\it powspec} task from
FTOOLS 5.1. The timing analysis PCA/{\it RXTE} data was performed in  13 -- 30 keV energy range 
using  the {\it event} mode.  The time resolution for this mode is 1.2$\times 10^{-4}$ s. We
generated power density spectra (PDS) in  0.1 -- 500 Hz frequency range
using light curves with $10^{-3}$-second time resolution. We subtracted the contribution due
to Poissonian noise and Very Large Event Window for all PDSs. We used 
QDP/PLT plotting package to model PDS.

%{\it
Previously, timing analysis of PCA/{\it RXTE} data for GX~3+1  observed on August 1999 (our R3 set), when 
the sources was in {\it lower banana}  phase,  was made by \cite{ooster01}.
%(LB) 
We investigated a timing behavior of GX~3+1 for our data set during all %spectral state
 transitions between {\it lower banana} %(LB) 
and {\it upper banana} %(UB) 
phases (see Fig.~\ref{PDS}). 
In particular, 
%as a function of source position in color-color diagram for RXTE data ($R1$, $R2$ sets in
%our terms). In island part of the color-color diagram (corresponding to the hardest energy
%spectra) 
the power spectrum of GX~3+1 consists of very low-frequency noise (VLFN, described by a power-law) 
and high-frequency noise [HFN, described by a cutoff-power-law, see \cite{klis05}].
%{klis95}].
% van der Klis 1995 

In 
%{\it lower banana} 
the LB 
phases 
(A $red$, 60022-01-13-01, MJD=52554; A $blue$, 60022-01-01-00, MJD=\ 51998) 
 the best-fit to the average power spectrum results in rms VLFN-component of 2\% %%%%%10.5\%
 (in the 0.1 -- 1 Hz range) 
described  by power-law  $\nu^{-\alpha_{LF}}$ 
with the index of  %$\alpha\sim$1.7, 
$\alpha_{LF}\sim$1.7,  
whereas  HFN  rms (in the 1 -- 100 Hz range)  has  4.7\% with $\alpha_{HF}\sim$1.0 and  $\nu_{cutoff}\sim$ 30 Hz. 
Generally %Remark here that 
the index of VLFN  $\alpha_{LF}$ gradually decreases from 1.7 to 1.4 toward to  
%{upper banana 
 UB. 
However in the vicinity of a transition point  
between   LB 
and UB 
(red histogram of panel B of Fig.~\ref{PDS}) $\alpha_{LF}$ jumps to 1.8 
(B $red$, 94307-05-01-00, MJD=55129)  and decreases again to 1.4 (B $blue$, 60022-01-11-03, MJD=52357). 
In general, 
%in the UB state 
the UB power spectra of GX~3+1 are dominated by the VLFN with the brake at 
%as a peaked noise component 
around 20 Hz at the lowest  $kT_e=2.4$~keV 
%of the UB 
(see $blue$ B point on the 
right-hand panel of Fig.~\ref{PDS}). %mostly brigh point of light curve. 
Specifically, 
during  UB
% {\it upper banana} (UB) 
($blue$ histogram of panels B of Fig. \ref{PDS}) %(B $red$, 94307-05-01-00, MJD=55129; B $blue$, 60022-01-11-02, 
%MJD=52394) 
one can see strong VLFN (rms=5.1$\pm$0.4\%, $\alpha_{LF}$=1.4$\pm$0.3, $\chi^2$=139 for 102 d.o.f; 
all parameter errors correspond to 1$\sigma$ confidence level) and 
HFN with rms=1.7$\pm$0.3\% and break frequency shifted from 30 Hz to 6 Hz. 
After 
%current {\it upper banana} %(UB) 
 UB  GX~3+1 again returns to 
%{\it lower banana} 
the LB  
%(
%see left bottom 
 showing 
similar to panel A properties of power density spectra (see panel C of Fig. \ref{PDS}).
% in this Figure. %But 
%On the panel C one can %would like to 
%see  some of 
%particular intermediate phases in addition to those  
%transitions 
%presented in panel A. 
%Namely, when GX 3+1 
%this atoll object becomes to 
%transits from 
%{\it upper banana} 
%the UB 
%state, firstly the HFN component 
%increases (from 1\% to 5\%; C $red$, 40023-01-04, %60022-01-04-03, MJD=52102)
%MJD=52404) along with VHFN decreasing (from 4\% to 2\%), finally 
%VHFN %can 
%absent at all during some of LB %low luminosity 
%intervals (C $blue$, 40425-01-01-06, MJD=51421).

Note that these components and their CD evolution are typical for
% the rest 
of atoll-sources in the banana state \citep{hasinger89} and caused by mass accretion rate change \citep{klis05}.
This  phase identification is supported by a combination of spectral (see Sect.~\ref{evolution}) 
and timing properties 
%(Sect.~\ref{transitions}) 
in an agreement with $atoll$--$Z$ scheme. %Note that the track of Fig.10 in whole reflects $long-term$ 
%evolution of GX~3+1, whereas $slow$ evolution thereof is presented as reciprocate along the track without visual 
%effect (not perpendicular to a track pass). Small scattering of points of said track in this CD is caused by 
%long-term evolution during some cycles with different amplitudes. We can suppose that in ideal case we would see 
%the parallel tracks, at least, of {\it upper banana} branch.

%We find  a similar  timing behavior of GX~3+1 for our data set during all %spectral state
%%%% BEGIN
%{\it
%It is worth to note that ASM/RXTE data timing analysis (1996 -- 2001) of GX~3+1 performed at 
%very low frequencies ($10^{-7}$ -- $5\times 10^{-5}$ Hz)  by \citet{reig03} shows rather moderate variability 
%(20\%$le$rms$le$30\%) of power spectra (0.9$\le\alpha\le$1.2) in this range.

While the aforementioned CD evolution of power spectra of GX~3+1 occurs on time scales from hours to days, 
we detected similar patterns for power spectrum evolution during LB -- UB transitions for both 
{\it faint phase} %low luminosity} 
and 
{\it bright phase }on luminosity %high luminosity} states 
during long-term variability within 14 years with  a quasi-periodic trend during six years.  
The  similarity of LB -- UB transitions which are independent of bright/faint phases on luminosity %low/high luminosity states 
indicates to  similar accretion configurations in all phases.

In a previous analysis of other RXTE data of GX~3+1 \citet{ooster01} %Oosterbroek et al. (2000) 
report VLFN and HFN values, in LB state, 
which are similar to our values in interval R3, with the exception 
of VLFN strength for which they report 7.5\% rms while we find 1.7\% rms.
% in that state. 
All of  the VLFN and 
HFN values of the analysis of EXOSAT data reported by~\cite{hasinger89} %Hasinger \& van der Klis (1989)
 for GX 3+1 agree with our results.
%}                  
%%%END

\subsection{Comparison of spectral and timing  characteristics {\it atoll} sources GX 3+1 and 4U 1728-34} 
%as a function of mass accretion rate  \label{disc}}

\subsubsection{Quasi-Constancy of the photon index}

The $atoll$ sources GX~3+1 and 4U~1728-34 demonstrate a similar behavior of the parameter $\Gamma$ vs 
mass accretion rate (or our COMPTB normalization), namely  the quasi-constancy of  the photon 
index $\Gamma\approx 2$ and 
 almost identical  long-term variations of ASM mean count rate  (see also ST11). %Fig.~\ref{outburst_index_temperature2}).
According to FT11  and ST11 this observational fact can presumably  indicate  that the energy release in the transition layer for  these 
two sources  is much higher than cooling flux of the soft (disk) photons. 

%    The relative variations of the COMPTB normalization  for GX 3+1 (from 0.04 to 0.15) and for  4U 1728-34
% (from 0.02 to 0.08)  are also similar.   Although variations of inferred electron temperature $kT_e$ 
% %as the best-fit parameter of COMPTB spectral component 
% is much smaller in the case of  GX 3+1 (from 2.5  to 4.5 %3.7 
% keV)  than that for 4U 1728-34  (from 2.5  to 15 keV).  
% Whereas variation of Comptonization fraction $f$  is a factor of 4 larger in the case of  GX 3+1 (from 0.1  to 0.9)  than that   for 4U 1728-34 (from 0.5  to 1).  
% % different for these while show different behavior of 
%%the electron temperature $T_e$ and COMPTB Normalization during state transition 

\subsubsection{A difference  of the electron temperature $kT_e$ ranges in GX 3+1 and 4U 1728-34}

A comparison of  the best-fit spectral parameters 
%evolution 
for  these two  {\it atoll} sources  shows that
a $slow$ variability is generally related to changes of COMPTB normalization, and  a $mild$
variability is mainly correlated with the electron temperature variations  (see Fig \ref{color_diagram}). 
Note that the ranges of disk and NS temperatures   
%(conditions) 
are similar  for both of these objects, namely  $kT_{BB}\simeq 0.6$~keV and $kT_s=$1.2 -- 1.7 keV respectively.   On the other hand  variations of the electron temperature  $kT_e$ are  quite different.
% for these objects. 
The electron temperature  $kT_e$ changes %varies 
in a wide range  $kT_e$=2.5 -- 15 keV  for 4U~1728-34, while for GX~3+1  $kT_e$ varies 
in a narrow range from 2.3 to 4.5 keV 
(see Figs.~\ref{T_e_vs_f_comp} %outburst_index_temperature1}
-- \ref{norm_T_e}). %outburst_index_temperature2}).
%It is possible that %well known 
%by this reason these atoll sources are quite different on CD track.
 The reason for this difference of temperature ranges is quite obvious.  While  4U~1728-34 shows an evolution
 from the extreme island state (EIS) 
 %through $island$ state (IS), to  lower left banana state (LLB), 
%traces the lower banana state (LB) and achieves 
to the upper banana state (UB) [see, \cite{disalvo2001} and ST11]  GX~3+1 demonstrates only 
a short LB -- UB track on CD (see Fig. \ref{T_e_vs_f_comp}). %outburst_index_temperature1}). 
%According to results of previous Sect.~4.1 
These ranges of CD states are related to 
the ranges of 
%the electron temperature 
$kT_e$. 
%of Compton cloud (TL). 
% of Comptonized zones in accretion flows.
%It is reasonably (possibly) to link these ranges of CD states with the ranges of the electron 
%temperature $kT_e$ of Comptonized zones in accretion flows. In the next section we will argue (discuss) this point in relation to 
%atoll source GX~3+1 based on the evolution of timing properties. Looking ahead, 
%Note also  that the evolutions of timing properties 
%for both 4U~1728-34 and GX~3+1 are similar and correspond to common states (LB -- UB) on CD.
%demonstrates larger luminosity amplitude of mild variability, than that for 
%GX~3+1. Specifically, for these objects: ${\Delta T^{4U}_e}$=2.5-15 keV and ${\Delta
%T^{GX}_e}$=2.3-3.7 keV.
 
\subsubsection{Comparison of spectral evolution as a function of the COMPTB normalization 
%(disk mass accretion rate)  
for GX 3+1 and 4U 1728-34}

We can also compare spectral parameter evolution for GX~3+1 and 4U~1728-34 using COMPTB
normalization because the distances to these sources are almost the same (see 
Table 5). Namely, for GX~3+1 the distance is in the range of 4.2$-$6.4 kpc~\citep{kk00},  
%  [van Paradijs (1978)] 
whereas for 4U~1728-34 it is 4.5 kpc~\citep{par78}.
%[Kuulkers \& van der Klis (2000)]. 
In Figure \ref{norm_T_e} %{outburst_index_temperature2}  %(bottom panels) 
we show 
%the
% measured 
a correlation of % $\Gamma$ 
COMPTB normalization presumably proportional to mass accretion rate) and   %COMPTB normalization (left), the Comptonization fraction $f$ (center) and 
the electron temperature $kT_e$ for these  two {\it atoll}
sources. %(right). 
%The last parameter is
%extremely suitable to compare spectral properties of different objects because of
%$kT_e$ is independent from source distance to Earth observer $D$. 
%As well as,
%parameter $kT_e$ is suitable to trace the source spectral states, particularly for
%NSs, because $kT_e$ decreases when sources move from the low state to the high state during mild
%variability as a result of a more efficient electron cooling by the increased seed
%photon supply. Moreover, the electron temperature $kT_e$ is a directly measurable
%quantity. 
GX~3+1 demonstrates  a wider  range of COMPTB
normalization (by  factor of 2   higher than that for  4U~1728-34) 
%(along vertical axis of Fig.~\ref{outburst_index_temperature2}), 
while 
%the electron temperature 
$kT_e$ varies 
only from 2.5 to 4.5 keV.
% (along horizontal axis of Fig.~\ref{outburst_index_temperature2}).
%varies , resting close to  3 keV.
%According to FT11 and  ST11 and the given investigation the electron
%temperature $kT_e$ for  atoll and Z-sources varies  from 2.5 to 25 keV. 
The common interval of $kT_e$ for GX~3+1 and 4U~1728-34 is in the range  2.5 -- 4.5 %3.7 
keV only and the  low limit of the electron temperature of 2.5 keV takes place
at the  peak luminosity for 4U~1728-34 (see ST11) and during increases in luminosity for GX~3+1, i.e. during  so called UB state (see 
%Figs.~\ref{lc_1998} and 
Fig.~\ref{T_e_vs_f_comp}). %outburst_index_temperature1}).
% of $mild$ variability.
%%% BEGIN
%{\it
%Thus we suggest  CD for both sources are released during LB -- UB states, 
%which possibly traced along this narrow $kT_e$ range (2.5 -- 4.5 keV).
%}
%%% END
 %Note also that $kT_e$, collected by FT11
 %for a number of  NS sources, is concentrated around  $kT_e\sim$~3 keV, 
%which, as we  show  further, indicates to the presence of $banana$ (B)  states for these sources.
%Possibly  this value of the electron temperature is related with crucial
%value for stable transfer of matter from the accretion disk onto NS through
%transition layer.

%\subsubsection{The Comptonization fraction parameter $f$ traces different ranges of states on color-color diagram}

\subsubsection{Correlation of illumination parameter  $f$ vs electron temperature $kT_e$ and its relation  with different states on color-color diagram}

One can see from Table 5 %the bottom panels of %Finally, middle of  Figure \ref{norm_T_e} %{outburst_index_temperature2} 
that  the range of illumination fraction of Compton cloud (TL)  $f$ is wider  for GX~3+1 
($0.1-0.9$)  than that for 4U 1728-34 ($0.5-1$). It can be related  to different illumination of the transition 
layer (TL) for these two sources. For 4U~1728-34 the solid angle viewed from NS changes by factor 2  whereas in  GX~3+1 that changes 
%the illumination of TL by photons coming from NS surface  and  thus the solid angle  changes 
by factor 4. 
However,  the photon index  $\Gamma\approx 2$
% is almost constant, around 2 
for these two sources which indicates that the energy 
release in the transition layer for  these two sources  is much higher than cooling flux of the disk photons 
(see FT11 and ST11 for details of  X-ray spectral formation in TL). 

\begin{figure}[ptbptbptb]
\includegraphics[scale=0.98,angle=0]{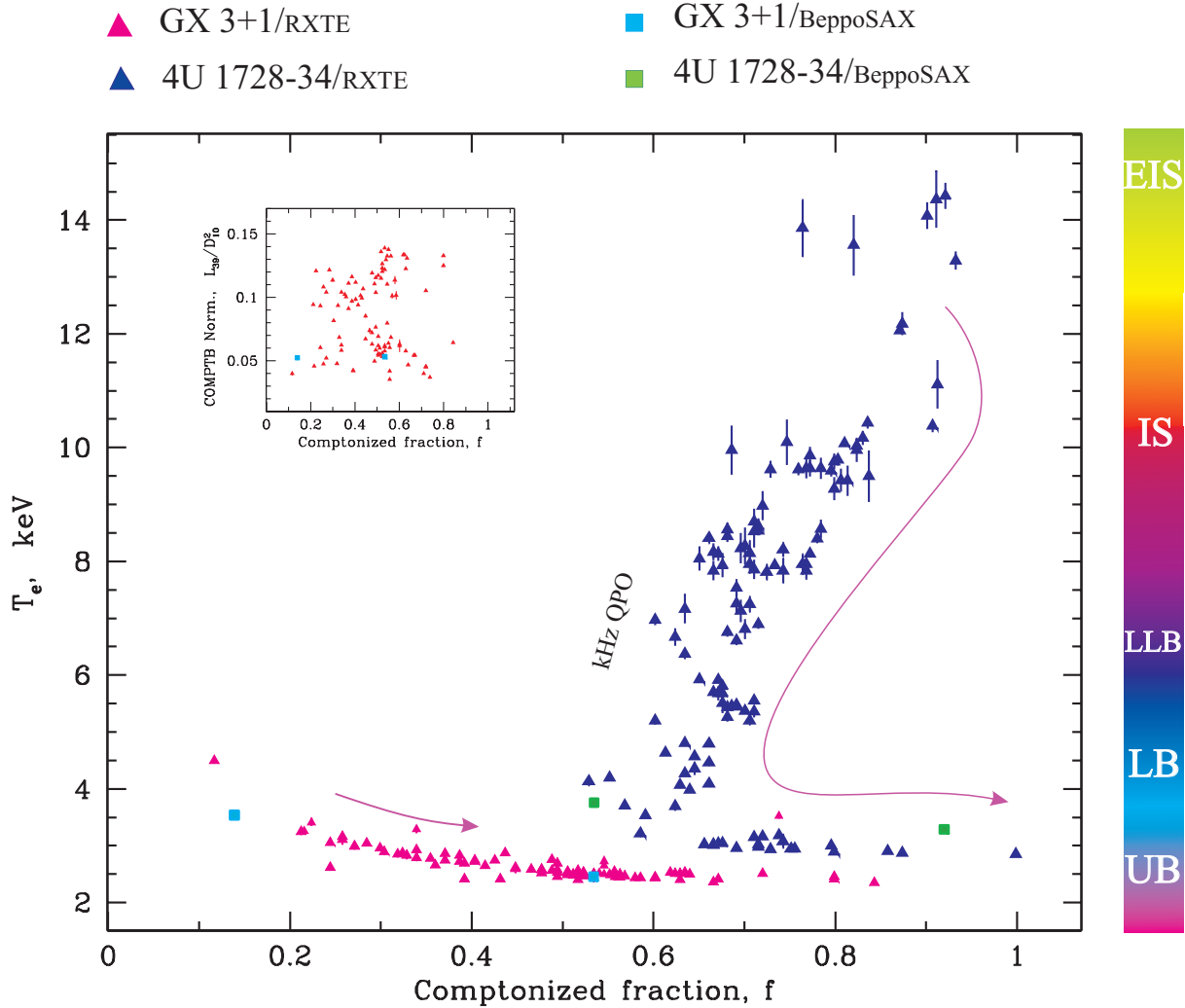}
\caption{
Electron temperature $T_e$ (in keV)  plotted versus illumination  fraction $f=A/(1+A)$ for atoll sources GX~3+1 and 
4U~1728-34 during $mild$  variability. {\it Pink/bright blue}  and {\it blue/green} points correspond to 
{\it RXTE}/{\it Beppo}SAX 
observations of GX~3+1 and 4U~1728-34 respectively. For GX~3+1 COMPTB normalization measured in $L^{soft}_{39}/D^2_{10}$ units 
versus illumination fraction $f$ is plotted in the {\it incorporated panel} ({\it top left})  
%using  our spectral model 
%$wabs*(blackbody+COMPTB+Gaussian)$ 
during long-term ($slow$) variability (see Table 4). The {\it bended arrows} are related to   an increase of mass accretion rate.
%  along the corresponding tracks. 
On the right-hand side
% of the  Figure 
we show  a sequence of CD states 
(EIS -- the extreme island state, 
IS --  island state,
LLB -- lower left banana state,
LB -- lower banana state and 
UB -- upper banana state) which  are listed according to the  standard atoll-Z scheme \citep{hasinger89}.
One can see that  
%which here matched by 
%the electron temperature 
$T_e$  is directly related with the sequence of CD states.
% with the digitalization (calibration) on left vertical axis. 
Along the track  for 4U~1728-34  we indicate   points of $T_e-f$  correlation where kHz QPOs are detected.   %indicated branch referring to the discussion in Sect.~4.2.6.
}
%\label{outburst_index_temperature1}
\label{T_e_vs_f_comp}
\end{figure}

In  Figure~\ref{T_e_vs_f_comp} %{outburst_index_temperature1}  
we present a plot $kT_e$
versus
% illumination fraction 
$f=A/(1+A)$ for atoll sources GX~3+1 and 4U~1728-34 
during  $mild$  variability. 
{\it Pink/bright blue}  and {\it blue/green} points correspond to {\it RXTE}/Beppo$SAX$ 
observations of GX~3+1 and 4U~1728-34 respectively. COMPTB normalization measured in $L^{soft}_{39}/D^2_{10}$ units  versus 
%Comptonized fraction 
$f$ is plotted in the { incorporated top left panel} 
%({\it top left}) 
%using  our spectral model 
%$wabs*(blackbody+COMPTB+Gaussian)$ 
for  long-term ($slow$) variability of GX~3+1 
(see Table 4 for details). 
%The {bended arrows } are made
%along  an increase of mass accretion rate direction.
% along the corresponding tracks. 
The { bended} arrows along the corresponding tracks
correspond to an  increase in mass accretion rate.   
On the right-hand side of Figure we also present  the sequence of CD states 
%(EIS -- the extreme island state, 
%IS --  island state,
%LLB -- lower left banana state,
%LB -- lower banana state and 
%UB -- upper banana state) 
which are listed according to the standard atoll-Z scheme  \citep{hasinger89}.
 Here we also show that the CD sequence  is definitely related to  the electron temperature $kT_e$.
 %  with the digitalization (calibration) on left vertical axis. 
 The diagram of $T_e$ versus $f$ 
demonstrates a  clear correlation $T_e$ and  $f$   while the diagram $N_{COM}$ versus $f$,  presented in the  incorporated panel 
of Fig.~\ref{T_e_vs_f_comp} %{outburst_index_temperature1}  
shows only chaotic scattering of points in a wide range of $f\sim$ 0.2 -- 0.9.  

Moreover, we find  two different track shapes  on diagram of $T_e$ versus $f$  for two atoll sources GX~3+1 and 4U~1728-34  and  their relations with 
%revealed  which 
%occupy definite diagram areas and form portions, corresponding to 
the standard sequence of CD states (Fig.~\ref{T_e_vs_f_comp}). %{outburst_index_temperature1}) . 
When  the  fraction $f$ increases, the electron temperature $T_e$ decreases 
monotonically from approximately 4.5 keV to $\sim$ 2.3 keV for GX~3+1, while 4U~1728-34 demonstrates more complicated behavior pattern.  At  EIS, with a decrease of $T_e$, the fraction $f$ slightly varies from 0.9 to 1.  When  $T_e$ further decreases from 12 keV to  4 keV,  $f$ decreases 
%become to decreases 
from 0.9 to 0.5.  Finally, during the LB-UB state transition $f$ increases from 0.5 to 1. 

Thus we show  that the evolution  CD states can be traced by the correlation between $T_e$ and $f$ too. 
Note that the tracks of $f-T_e$ on this diagram resemble
%  very similar  reminiscent (resemble) 
the  atoll tracks on the standard color-color diagram~\citep{schulz89}.

\section{Conclusions \label{summary}} 

We present our analysis  of the spectral properties observed in X-rays from 
the 
%{\it
%burstimg atoll source
%}
neutron star  X-ray binary  GX~3+1 during long-term transitions
between the $faint$ phase %low state  
and the $bright$ phase % high state 
%{it
superimposed by short-term transitions between {\it lower banana} and {\it upper banana} states. 
%}
We analyze all %a number of  
transition episodes 
for this source  observed with {\it Beppo}SAX  and 
{\it RXTE}.
% satellites.  
For our  analysis we apply   the good spectral coverage and  resolution of  $Beppo$SAX detectors 
from 0.1 to 200 keV along  with $extensive$  $RXTE$ coverage in the energy range from 3 to 50 keV.    

\begin{figure}[ptbptbptb]
\includegraphics[scale=0.9,angle=0]{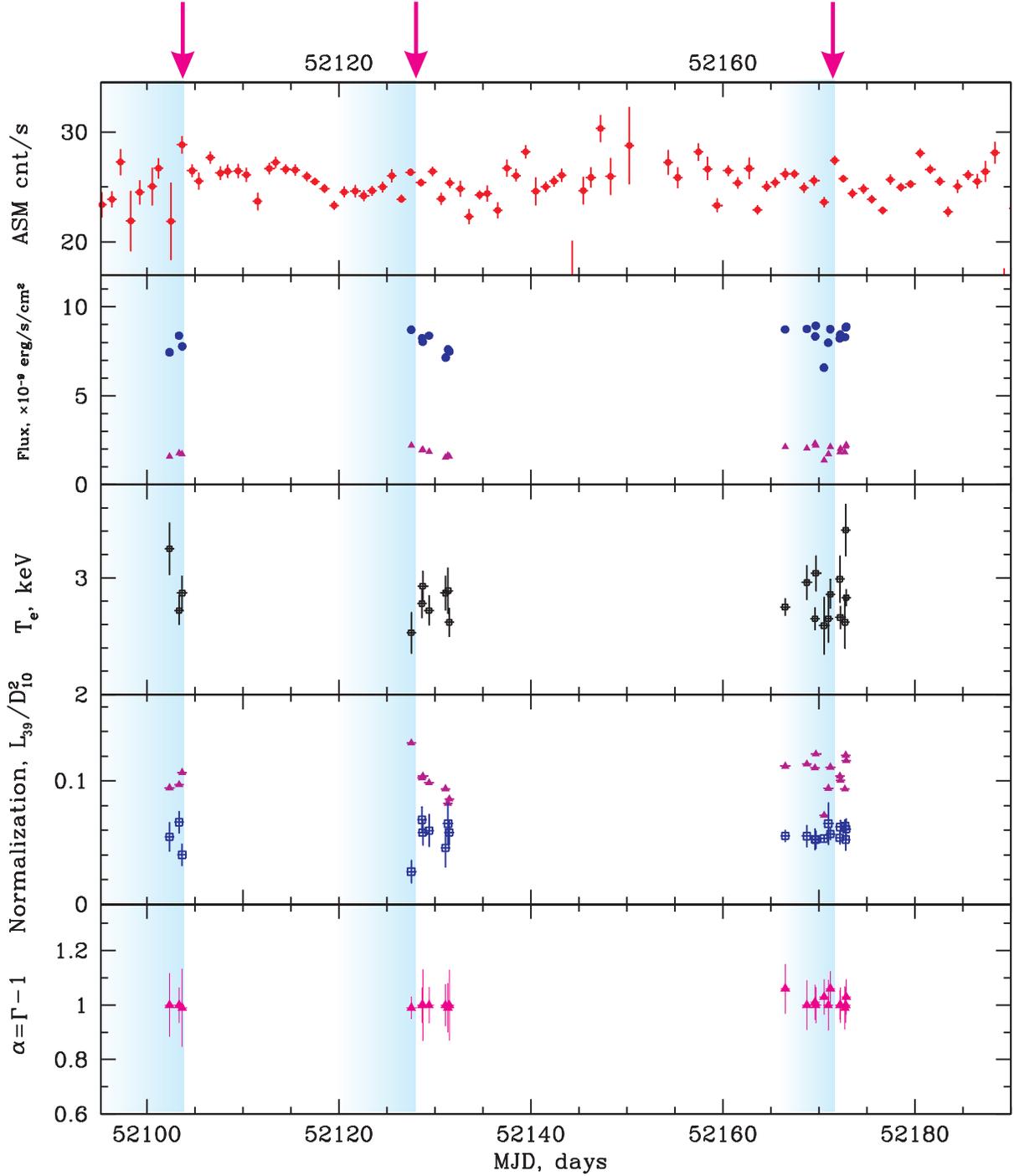}
\caption{{\it From Top to Bottom:}
Evolutions of the ASM/{\it RXTE} count rate, model flux in 3-10 keV  and 10-60 keV energy ranges 
({\it blue and crimson} points respectively), electron temperature $T_e$ in keV, 
 $COMPTB$ and $blackbody$ normalizations 
({\it crimson, blue} points respectively)  and  spectral index $\alpha=\Gamma-1$ during 2001$-$2002 outburst transitions set ({\it R4}). 
The rising phases  of the $mild$ transition are marked with blue vertical strips. 
The peak burst times are indicated by the arrows on the top of the plot.
% to point out demonstration (illustration) of $mild$ variability.
}
\label{lc_1998}
\end{figure}

\begin{figure}[ptbptbptb]
\includegraphics[scale=0.9,angle=0]{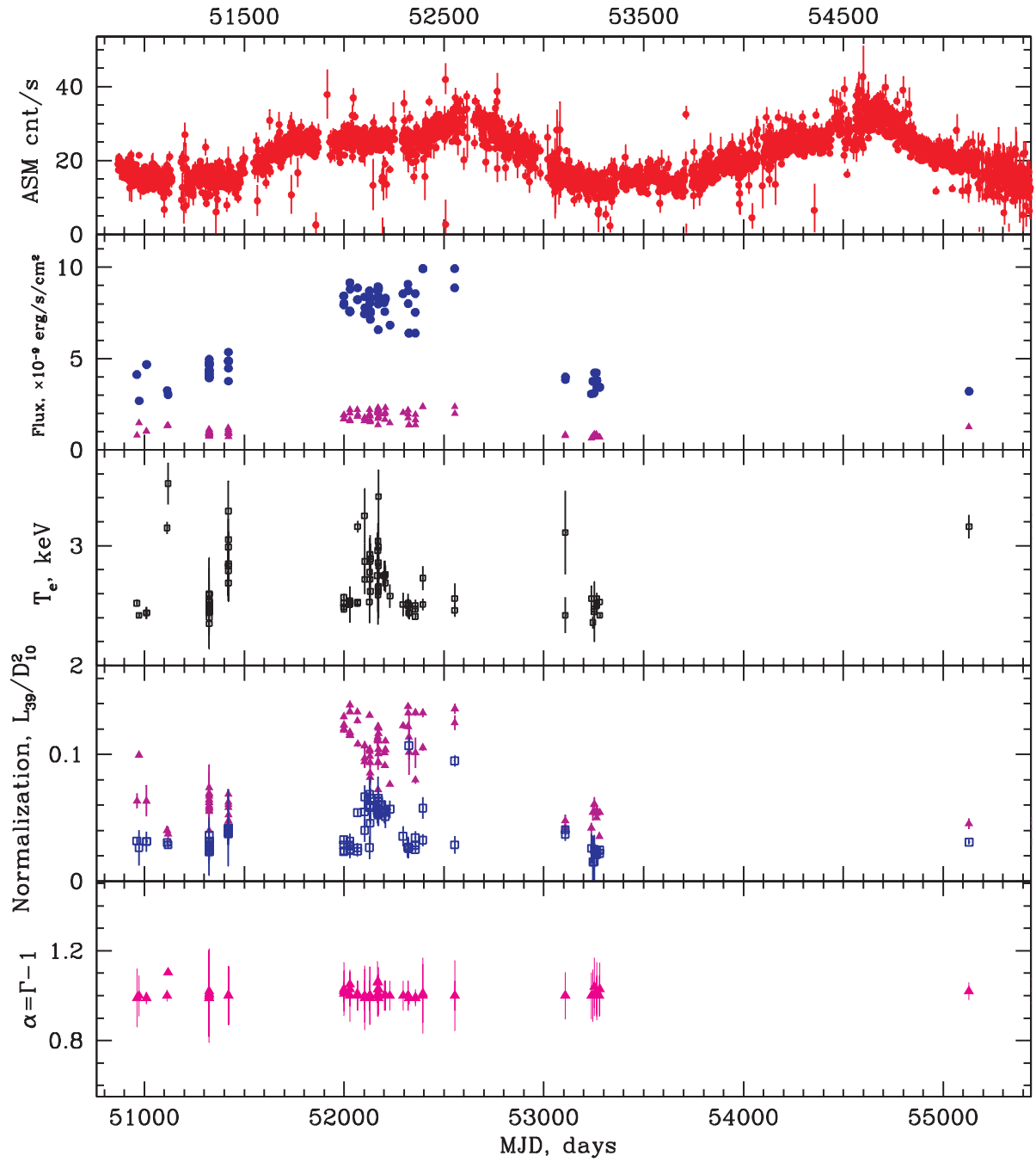}
\caption{Same to  that presented in Fig. \ref{lc_1998} but  for all the {\it RXTE} 
%2000 outburst transition 
sets ({\it R1 - R7}) to  demonstrate 
%(illustration) 
$slow~(long-term)$ variability.
}
\label{evolution_lc_all}
\end{figure}

\begin{figure}[ptbptbptb]
\includegraphics[scale=0.92,angle=0]{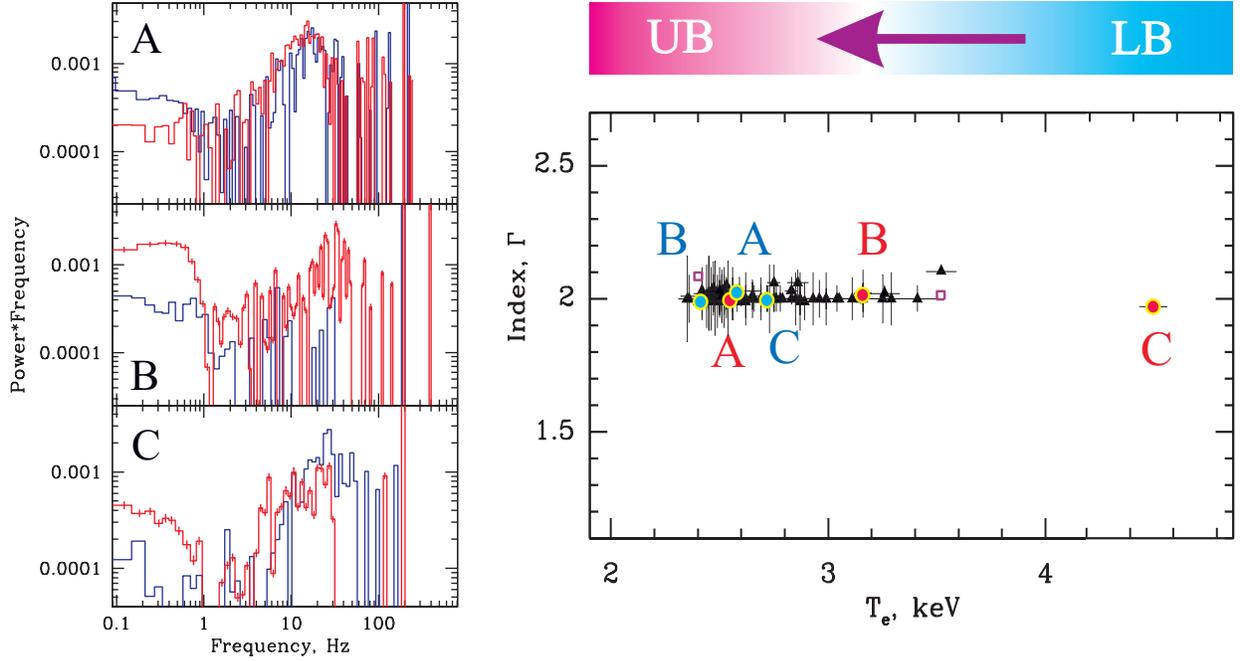}
\caption{
{\it Left column:} 
PDSs presented  in terms of   $\nu P(\nu)$  for 13 -- 30 keV energy band correspond to  {\it lower banana} %(LB) %(panels A, C) 
and 
{\it upper banana} %(UB) %(panel B) % and additional phases of {\it lower banana} states (panel ``C'')
states of GX~3+1 
 and related 
to different electron temperatures indicated by points A, B and C on the {\it right panel}  of the Figure.
%diagram.
 %representation. %$left$ column. 
The strong noise component HFN %(rms$\sim$5\%)
 seen in  the 1 -- 50 Hz range and relatively weak VLFN %(rms$\sim$2\%), 
%(described  by power-law $\nu^{-\alpha}$) %with the index of  $\alpha_{VLFN}$, 
at the frequences below $\sim$ 1 Hz are present  before %(A $red$, 60022-01-13-01, MJD=52554; A $blue$, 60022-01-01-00, MJD=51998)  
and after transition (see panels A and C) %of GX~3+1 
from LB to UB. %{\it lower banana} state (see panels A, C) to {\it upper banana} state.   
In  UB %{\it upper banana} state
 (panel B, $blue$ histogram) the power spectra of GX~3+1 are dominated by the VLFN with a break at about %as a peaked noise component around 
20 Hz. 
%The power-law index  $\alpha$ of VLFN gradually decreases from 1.7 to 1.4 during LB$\to$UB transition
%%%%toward to the UB %{\it upper banana} state
%and vice versa, wherein %during UB$\to$LB transition %to UB 
%%(red histogram of panel B)
%% $\alpha_{VLFN}$
%there is a point where $\alpha$ jumps to 1.8 with following decreasing to 1.4 at the vicinity of LB ($blue$ histogram of panel B). %see details in the text). 
%(B $red$, 94307-05-01-00, MJD=55129)  and decreases again to 1.4 (B $blue$, 60022-01-11-02, MJD=52394). 
%Some of particular intermediate phases are demonstrated in panel C (in addition to panel A): % of LB state of GX~3+1. 
% when GX~3+1 becomes to transit from UB % {\it upper banana} (UB) states
%, firstly the HFN component 
%increases (from 1\% to 5\%; C $red$, 60022-01-04-03, MJD=52102) along with VHFN decreasing (from 4\% to 2\%), finally 
%VHFN can absent at all during low luminosity intervals (C $blue$, 40425-01-01-06, MJD=51421).
{\it Right panel:} Photon index $\Gamma$ plotted versus %COMPTB normalization  
%$N_{COMPTB}$ measured in $L^{soft}_{39}/D^2_{10}$ units (panel $a$),  $\Gamma$ versus 
electron temperature $T_e$ (in keV) %(panel $b$) 
%and  versus 
%Comptonized fraction $f=A/(1+A)$ (panel $c$), 
%COMPTB normalization versus Comptonized fraction $f$ (panel $d$) 
using  our spectral model 
$wabs*(blackbody+COMPTB+Gaussian$) during spectral  transitions (see Table 4). 
 Violet and   black points correspond to {\it Beppo}SAX  and {\it RXTE} observations of GX~3+1 
%in {\it left/right} panels 
respectively. The violet arrow in {\it top part of the right panel} points the direction of LB$\to$UB transition related to a decrease of $T_e$.
}
\label{PDS}
\end{figure}

We show that the X-ray broad-band energy spectra during all spectral states can be adequately fitted by   the combination of  a {\it Blackbody},  a Comptonized
%({\it COMPTB}) 
and  a  {\it Gaussian} components. 
 We  also show  that photon index $\Gamma$ of the best-fit Comptonized component  in GX~3+1 is 
almost constant, with a value of  2 
%(see Fig. \ref{outburst_index_temperature1}) 
and consequently is almost 
 independent of   {\it COMPTB} normalization $L^{soft}_{39}/D^2_{10}$  which is presumably proportional to  mass accretion 
rate $\dot m$
% and plasma temperature of 
%Compton cloud $T_e$  
(see Figs. \ref{lc_1998} -- \ref{evolution_lc_all}, \ref{PDS}).
% \ref{outburst_index_temperature1}
% Note the soft (disk) photon luminosity 
%$L_{39}^{soft}$ is units $10^{39}$ erg s$^{-1}$ and distance to the source $D_{10}$ is units of $10$ kpc. 
 We  should remind a reader  that this index stability  has recently been  suggested using  X-ray observations of quite a few  of other  NS sources.
%%% BEGIN
Namely   atoll sources:  X 1658-298, GS 1826-238, 1E 1724-3045  and also 
 Z-sources: Cyg~X-2, Sco~X-1, GX~17+2, GX~340+0, GX 349+2
%%% END
% Cyg X-2, Sco X-1, GX 17+2,GX 340+0, GX 3+1, GX 349+2, X 1658-298, GS 1826-238, 1E 1724-3045 
 were  observed by {\it Beppo}SAX at different spectral  states  and also  atoll source 4U~1728-34 
observed by {\it Beppo}SAX  and {\it RXTE} [see details in  FT11 and ST11 respectively].

A wide variation of  parameter  $f=0.1-0.9$, obtained in the framework of our spectral model,   points out 
%gives us  a strong  argument  
a significant  variation of the illumination of  Comptonization region (transition layer)  by  X-ray soft photons
%disk emission 
%in Compton cloud 
 in  GX~3+1.    
%  The use of th ({\it COMPTB}) normalization, which is 

 Using {\it Beppo}SAX  observations we find  that there are two sources of blackbody emission, 
one is presumably related to   the accretion disk  and another one is related  to the  NS surface for which  
temperatures of soft photons are about 0.7 keV and 1.3 keV, respectively.  

%{\it
We demonstrate that our analysis of X-ray spectral and timing properties in  atoll source GX~3+1 allows us  to 
distinguish  between $mild$ and long-term variabilities, and link them with LB -- UB state transitions and transitions 
between $bright$ and $faint$ phases in luminosity, respectively. In this way we described $mild$ flux variability 
between LB and UB states on time scale of hours -- days  in terms of two  basic spectral parameters, 
the electron temperature $kT_e$ and illumination fraction $f$.
%as brighter -- fainter phase % high -- low state 
%transitions and LB -- UB transitions, respectively, 
%which are associated with COMPTB normalization and the electron temperature of Comptonized layers changings, correspondingly.
We argue %This model allow to argue %Spectral analysis using this model provides 
that $kT_e$  monotonically increases from 2.3 keV to 4.5 keV when GX~3+1 makes a transition from UB  state 
%{\it upper banana} 
to  LB state.
%{\it lower banana}.
%  presumably related  to  small  flux variabilities on time scale of hours.  
%Specifically, we
We also find,
% at least,
 two noise components (VLFN \& HFN) %demonstrate 
and their evolutiona 
%(alternative behavior) 
during LB -- UB transitions: %, associated with low flux variability on time of hours, 
the X-ray power spectra (PDS) in UB 
%{\it upper banana} state  
are dominated by %a very low frequency noise 
very low frequency noise (VLFN) with 
%peaked noise component 
the break  around 20 Hz, whereas in LB
%{\it lower banana} 
%state 
the PDSs  are dominated by a high frequency noise 
(HFN) in 1 -- 100 Hz range and accompanied by reduced VLFN below $\sim$ 1 Hz.
%}
%%%%% END

We demonstrate  that
 the photon index $\Gamma\sim 2$ is approximately  constant  when  the source moves from the faint phase %low state 
 to the bright phase %high state 
%(see  Figs.~\ref{lc_1998}-\ref{evolution_lc_all}, \ref{PDS})
%\ref{outburst_index_temperature1} and \ref{outburst_index_temperature2}) 
%{\it
and as well as  during local transitions from {\it lower banana} %(LB)  
to {\it upper banana}. %(UB) states (see  Figs.~\ref{lc_1998} -- \ref{evolution_lc_all}, \ref{PDS}).
%} 
In ST11 we presented   strong theoretical arguments  that the dominance of the energy release in the transition layer (TL) 
%  thermal Comptonized component formed in the 
%transition layers 
 with respect to the soft  flux coming from the accretion disk,  $Q_{disk}/Q_{cor}\ll1$  leads
to almost   constant photon  index $\Gamma\approx2$.

 Thus we argue that {\it the stability of this index   is an intrinsic signature of  atoll  sources while in BHs 
the index monotonically changes with mass accretion rate and ultimately saturates} (see ST09 and ST11) ).
% It is worth to remind a reader  the index correlation vs mass accretion rate for a number of 
% BH sources and how  the index depends on mass accretion rate in NSs GX~3+1 and 4U 1728-34. 
Photon indices  of BH candidates  (GRS~1915+105, GX~339-4, SS~433, H1743-322,  4U 1543-47, Cyg X-1, XTE J1550-564 and GRO~J1655-40)   show  clear 
correlation with mass accretion rate $\dot m$ 
%or with soft photon normalization $L_{39}^{soft}/D^2_{10}$ which is 
%proportional to $\dot m$ 
[ST09, TS09 and \citet{ST10}].   
This correlation  is accompanied by an index saturation when  $\dot m$ exceeds a certain level.  
%in comparison to $atoll$ NS source (GX~3+1) sample 
 The behaviors of the index for  GX~3+1 and 4U 1728-34  are clearly  different 
  from that for the sample of  BHC  sources.   The photon index $\Gamma\approx2$
  % independently of any model parameter 
  while mass accretion rate changes by factor 4.
  
We acknowledge discussion and editing  of the paper content with Chris Shrader.
%{\it
We are very grateful to the referee whose constructive suggestions help us to improve the paper quality.
%}

\newpage
\begin{deluxetable}{ccccc}
%%%%%\rotate
\tablewidth{0in}
\tabletypesize{\scriptsize}
%  \begin{center}
    \tablecaption{The list of $Beppo$SAX observations of GX~3+1  used in our analysis.}
    \renewcommand{\arraystretch}{1.2}
%    \begin{tabular}[h]
%      \hline
\tablehead{
Obs. ID& Start time (UT)  & End time (UT) &MJD interval & CD state}
%Satellite&Obs. ID& Start time (UT)  & End time (UT)}
%%%%%Obs.  &ID           & time (UT)& time (UT)& of state& }
\startdata
20603001& 1999 Feb. 28 11:02:15 & 1999 Feb. 30 09:14:15 &51237.4-51238.9 & {\it upper banana}\\
20835001& 1999 Aug. 30 18:33:08 & 1999 Aug. 31 04:54:32 &51420.8-51421.9$^1$& {\it lower banana}\\
%20889003& 1999 Aug. 19 02:01:32 & 1999 Aug. 20 04:54:32 &51409.1-51410.2$^1$& \citet{piraino00}\\
      \enddata
%      \hline
%      \end{tabular}
   \label{tab:table}
% \end{center}
Reference. 
(1) \citet{ooster01} 
%Piraino et al., (2000)
\end{deluxetable}

\newpage
\begin{deluxetable}{lllllc}
%\rotate
\tablewidth{0in}
\tabletypesize{\scriptsize}
%  \begin{center}
    \tablecaption{The list of {\it RXTE} observation  groups  of GX~3+1}
    \renewcommand{\arraystretch}{1.2}
%    \begin{tabular}[h]
%      \hline
\tablehead{Number of set  & Dates, MJD & RXTE Proposal ID&  Dates UT & Rem. & Phase on \\
                          &            &                 &           &      &  lumonosity}
 \startdata
R1  &    50365       & 10069        & Oct. 9 03:36:00 -- 04:08:00, 1996 &          & $faint$ \\
R2  &    50962-51118 & 30042, 30048 & May 29 -- Nov. 1, 1998            &          & $faint$ \\
R3  &    51324-51445 & 40023, 40425$^1$ & May 26 -- Sept. 24, 1999      & $Beppo$SAX & $faint$\\
R4  &    51998-52554 & 60022        & March 30, 2001 -- Oct. 7, 2002    &          & $bright$ \\
R5  &    52881       & 80105        & Aug. 30 03:53:36 -- 06:29:13, 2003&          &  $faint$\\
R6  &    53108-53280 & 90022        & Apr. 13 -- Oct. 2, 2004           &          &  $faint$\\
R7  &    55440.6-55440.8 & 94307    & Sept. 1 15:00:32 -- 21:07:58, 2010&          &  $faint$ \\
      \hline
      \enddata
%      \hline
%      \end{tabular}
    \label{tab:par_bbody}
%  \end{center}
References:
(1) \citet{ooster01} 
%Strohmayer et al. 1996; 
%(2) Ford \& van der Klis (1998);
%(3) van Straaten et al. (2002); 
%(4) Di Salvo et al. (2001); 
%(5) Mendez, van der Klis \& Ford (2001); 
%(6) Migliari, van der Klis \& Fender (2003); 
%(7) Jonker, Mendez \& van der Klis (2000);
%(8) TS05
\end{deluxetable}

%begin{deluxetable}{cccc}
%%%%%\rotate
%\tablewidth{0in}
%\tabletypesize{\scriptsize}
%  \begin{center}
 %   \tablecaption{The list of $Beppo$SAX observations of 4U~1728-34  used in analysis.}
  %  \renewcommand{\arraystretch}{1.2}
%    \begin{tabular}[h]
%      \hline
%\tablehead{
%Obs. ID& Start time (UT)  & End time (UT) &MJD interval}
%Satellite&Obs. ID& Start time (UT)  & End time (UT)}
%%%%%Obs.  &ID           & time (UT)& time (UT)& of state& }
%\startdata
%20674001& 1998 Aug. 23 19:15:27 & 1998 Aug. 24 09:14:15 &51048.8-51049.4$^1$ \\
%20889003& 1999 Aug. 19 02:01:32 & 1999 Aug. 20 04:54:32 &51409.1-51410.2$^2$\\
%20889003& 1999 Aug. 19 02:01:32 & 1999 Aug. 20 04:54:32 &51409.1-51410.2$^1$& \cite{piraino00}\\
 %     \enddata
%      \hline
%      \end{tabular}
%   \label{tab:table}
% \end{center}
%Reference
%$(1)$ \cite{disalvo2000a}, $(2)$ \cite{piraino00} 
%Piraino et al., (2000)
%\end{deluxetable}

\newpage
\bigskip
\begin{deluxetable}{cccccccccccccc}
%\begin{deluxetable}{ccccccccccccccc}
\rotate
\tablewidth{0in}
\tabletypesize{\scriptsize}
%  \begin{center}
    \tablecaption{Best-fit parameters of spectral analysis of $Beppo$SAX 
observations of GX~3+1 in 0.3-50~keV energy range$^{\dagger}$.
Parameter errors correspond to 1$\sigma$ confidence level.}
%\vspace{1em}
    \renewcommand{\arraystretch}{1.2}
%    \begin{tabular}[h]
%      \hline
%ID               & day  & & &   $\Gamma-1$          &           &$L_{39}/d^2_{10}$& keV & keV   &  &  keV        &  & & & }
 \tablehead
{Observational & MJD, & $T_{BB}$ & $N_{BB}^{\dagger\dagger}$ &$T_s$ & $\alpha=$  & $T_e,$ & $\log(A)$ & N$_{COM}^{\dagger\dagger}$ &  E$_{line}$,&   $N_{line}^{\dagger\dagger}$ &  $\chi^2_{red}$ (d.o.f.)\\
ID             & day  & keV         &                           & keV  &$\Gamma-1$  & keV    &                                  &                                   &   keV       &                               &                        }
 \startdata%   id     MJD      kT_Bbody     N_bb     [kT_s]    alf       T_e       log    norm_COMPTB E_line[Sigma_l] N_line Xi_2(dof)  Flux3-10 Fl10-60
20603001   &        51237.5 & 0.47(3)& 2.65(2) & 1.30(3)& 0.99(7) & 3.68(5) & 0.09(4) & 4.18(3) & 7.4(1) &  0.55(4)& 1.08(457)\\
20835001   &        51420.8 & 0.62(5)& 1.61(1) & 1.21(5)& 1.07(4) & 2.4(2) & -0.32(8) & 3.56(2) & 6.0(1) &  0.43(4)& 1.16(445)\\     
      \enddata%     \hline
%      \end{tabular}
    \label{tab:fit_table}
%  \end{center}
$^\dagger$ The spectral model is  $wabs*(blackbody + COMPTB + Gaussian)$,
normalization parameters of $blackbody$ and $COMPTB$ components are in units of 
$L_{37}^{soft}/d^2_{10}$ 
%$erg/s/kpc^2$, 
where $L_{37}^{soft}$ is the soft photon  luminosity in units of 10$^{37}$ erg/s, 
$d_{10}$ is the distance to the source in units of 10 kpc 
and $Gaussian$ component is in units of $10^{-2}\times total~~photons$ $cm^{-2}s^{-1}$ in line.
$^{\dagger\dagger}$ Gaussian component is in units of $10^{-2}$ ? total photons cm$^{?2}$ s${?1}$ in line.
%$^{\dagger\dagger\dagger}$when parameter $\log(A)\gg1$, it is fixed to a value 1.0 (see comments in the text). 
%$\sigma_{line2}$ of Gaussian2 component is fixed to a value 0.01 keV (see comments in the text), 
%$^{\dagger\dagger\dagger\dagger}$spectral fluxes (F$_1$/F$_2$) in the (3 -- 10)/(10 -- 60) keV energy ranges, correspondingly, 
%in units of $\times 10^{-9}$ ergs/s/cm$^2$.
\end{deluxetable}

\newpage
\bigskip
\begin{deluxetable}{lcccccccccccccc}
%\begin{deluxetable}{cccccccccccccc}
\rotate
\tablewidth{0in}
\tabletypesize{\scriptsize}
%  \begin{center}
    \tablecaption{The best-fit parameters of spectral analysis of PCA/{\it RXTE} 
observations of GX~3+1 in 3-50~keV energy range$^{\dagger}$.
Parameter errors correspond to 1$\sigma$ confidence level.}
%\vspace{1em}
    \renewcommand{\arraystretch}{1.2}
%    \begin{tabular}[h]
%      \hline
%ID               & day  & & &   $\Gamma-1$          &           &$L_{39}/d^2_{10}$& keV & keV   &  &  keV        &  & & & }
 \tablehead
{Observational & MJD, & $\alpha=$  & $T_e,$ & log(A) & N$_{COM}^{\dagger\dagger}$ & $T_s$, & $N_{Bbody}^{\dagger\dagger}$ & E$_{line}$,& $\sigma_{line},$ & $N_{line}^{\dagger\dagger}$ &  $\chi^2_{red}$ (d.o.f.)& F$_1$/F$_2^{\dagger\dagger\dagger}$ \\
ID             & day  & $\Gamma-1$ & keV    &                           &                                      & keV    &                               &  keV       &       keV              &                    &                         &                                                & }
 \startdata%   id     MJD    alf     T_e       log     norm_COMPTB    kT_bb     N_bb    E_line      N_line   Xi_2(dof)  Flux3-10 Fl10-60
10069-03-01-00 &  50365.172 & 1.00(8) & 2.40(2) &  0.03(1) & 10.88(2)  &  1.10(8)  & 2.74(5) & 6.53(3) &  0.58(5) & 0.95(4) &1.1(67) &6.79/1.28 \\
30042-04-01-00 &  50962.598 & 1.0(1) & 2.52(1) & -0.04(3) & 6.3(1)  &  1.65(5) & 3.17(3) & 6.51(2) &  0.5(1) & 0.46(2) &1.50(67) &4.13/0.80 \\
30042-04-02-00 &  50973.668 & 1.02(7)& 2.42(1) & -0.1(1)  & 9.9(1)  &  1.17(4) & 2.6(1)  & 6.5(1)   &  0.5(1) & 0.37(8)&1.12(67) &2.69/1.49 \\
30048-01-01-00 &  51011.138 & 0.99(2)& 2.44(2) &  0.18(8) & 6.3(2)  &  1.45(8) & 3.12(8) & 6.42(8) &  0.6(1)  & 0.7(1)  &0.82(67) &4.68/1.03 \\
30042-04-03-00 &  51113.941 & 1.01(2)& 3.15(2) &  0.39(5) & 4.00(6) &  1.5(1)  & 3.05(2) & 6.53(1) &  0.50(8) & 0.42(1) &0.87(67) &3.26/1.31 \\
30042-04-03-01 &  51114.207 & 1.03(3)& 3.17(2) &  0.41(5) & 4.07(6) &  1.5(1)  & 3.07(2) & 6.58(1) &  0.52(7) & 0.44(1) &0.91(67) &3.62/1.56 \\
30042-04-04-00 &  51118.739 & 1.1(1) & 3.52(8) &  0.45(6) & 3.7(1)  &  1.7(5)  & 2.86(3) & 6.51(4) &  0.6(2) & 0.47(2) &1.18(67) &3.03/1.34 \\
40023-01-01-00 &  51324.737 & 1.0(3) & 2.48(2) &  0.02(1) & 5.98(8) &  1.6(2) & 2.62(8)  & 6.43(9) &  0.67(9) & 0.9(1)  &1.35(67) &4.68/1.03 \\
40023-01-01-01 &  51325.310 & 1.0(2) & 2.40(1) &  0.23(3) & 5.78(6) &  1.6(2) & 2.64(9) & 6.43(5) &  0.67(8)  & 0.73(7)  &0.75(67) &4.15/0.88 \\
40023-01-01-02 &  51325.542 & 1.(1)  & 2.46(1) & -0.01(1) & 5.85(4) &  1.6(2) & 2.38(8) & 6.45(5) &  0.62(6) & 0.73(8)  &1.26(67) &4.06/0.77  \\
40023-01-03-00 &  51390.248 & 1.0(1) & 2.54(1) &  0.01(1) & 6.20(6) &  1.56(8) & 2.6(1) & 6.49(7) &  0.60(7) & 0.7(1)  &0.94(67) &4.33/0.88  \\
40023-01-02-00 &  51398.243 & 0.9(2) & 2.53(1) &  0.04(2) & 5.61(5) &  1.4(1) & 2.68(9) & 6.45(5) &  0.64(6) & 0.76(9)  &1.01(67) &3.99/0.81  \\
40023-01-02-01 &  51398.188 & 0.9(1) & 2.50(2) &  0.02(2) & 5.50(5) &  1.5(1) & 2.78(7) & 6.45(5) &  0.48(8) & 0.61(9)  &1.01(67) &3.93/0.76  \\
40023-01-02-02 &  51399.644 & 0.9(1) & 2.49(8) &  0.01(1) & 6.93(8) &  1.2(1) & 2.47(5) & 6.46(9) &  0.69(9) & 0.96(8)  &0.53(67) &4.74/0.94  \\
40023-01-02-03 &  51400.131 & 1.01(6)& 2.51(5) &  0.06(2) & 6.19(7) &  1.2(1) & 3.16(4) & 6.38(5) &  0.8(1)  & 1.09(6)  &0.71(67) &4.34/0.86  \\
40023-01-02-04 &  51400.192 & 1.01(9)& 2.4(1)  &  0.06(2) & 5.77(6) &  1.18(3)& 2.83(5) & 6.45(9) &  0.52(5) & 0.61(6)  &0.85(67) &3.97/0.75  \\
40023-01-02-05 &  51400.259 & 1.0(1) & 2.6(1)  & -0.09(5) & 6.72(6) &  1.25(2) & 2.9(1) & 6.50(7) &  0.46(7) & 0.63(9)  &0.77(67) &4.64/0.93  \\
40023-01-02-06 &  51400.325 & 1.0(1) & 2.45(7) &  0.18(9) & 6.19(5) &  1.25(8) & 3.04(8)& 6.46(8) &  0.41(6) & 0.45(9)  &0.92(67) &4.26/0.92  \\
40023-01-02-07 &  51400.723 & 1.0(1) & 2.45(1) &  0.11(2) & 6.85(4) &  1.20(8) & 2.38(7) & 6.45(8) &  0.47(5)& 0.45(8)  &0.92(67) &4.70/0.98  \\
40023-01-04-00 &  51404.304 & 0.99(1)& 4.50(4) & -0.9(1)  & 3.99(9) &  1.82(3) & 3.63(9) & 4.55(5) &  0.50(7) & 0.70(6) &1.70(67) &4.16/0.94  \\
40023-01-03-01 &  51405.307 & 1.(1)  & 2.48(1) &  0.09(2) & 6.38(6) &  1.20(8) & 2.32(9) & 6.55(8) &  0.56(9) & 0.60(8) &1.28(67) &4.39/0.92  \\
40425-01-01-00 &  51420.779 & 1.0(1) & 3.05(7) &  -0.5(5) & 6.05(8) &  1.59(8) & 4.01(6) & 6.5(1)  &  0.6(1)  & 1.05(4) &0.82(67) &4.84/0.97  \\
40425-01-01-01 &  51420.846 & 1.09(5)& 2.83(9) &  -0.31(7) & 6.85(7) &  1.59(9)& 4.09(9) & 6.5(1)  &  0.62(9) & 0.87(9) &1.05(67) &5.36/1.21  \\
40425-01-01-02 &  51421.018 & 1.05(1)& 2.8(1)  & -0.29(3) & 6.23(2) &  1.59(5) & 3.74(9) & 6.5(1)  &  0.61(8) & 0.85(2) &1.34(67) &4.88/1.10  \\
40425-01-01-03 &  51421.084 & 1.01(1) & 3.29(9) &  -0.29(3) & 5.83(5)&  1.59(8)& 4.19(6) & 6.5(1)  &  0.65(5) & 0.86(3) &1.30(67) &4.87/0.98  \\
40425-01-01-04 &  51421.151 & 1.0(1)  & 2.85(8) & -0.33(6) & 4.77(4) &  1.55(7)& 3.91(3) & 6.38(2) &  0.58(7) & 0.99(2) &1.15(67) &4.87/0.98  \\
40425-01-01-05 &  51421.218 & 1.08(9) & 2.99(5) & -0.43(3) & 5.22(5) &  1.49(8) & 4.11(2) & 6.38(4) &  0.59(4) & 1.11(8) &1.39(67) &4.47/0.88  \\
40425-01-01-06 &  51421.285 & 1.02(8) & 2.69(6) & -0.19(6) & 4.24(2) &  1.35(7) & 3.79(3) & 6.38(3) &  0.62(8) & 0.87(6) &1.36(67) &3.77/0.74  \\
40023-01-05-00 &  51445.320 & 0.99(3) & 2.59(9) & -0.06(4) & 7.38(7) &  1.24(8) & 2.7(1)  & 6.47(8) &  0.46(4) & 0.62(4) &0.84(67) &4.96/1.03  \\
40023-01-03-01 &  51445.599 & 1.0(1)  & 2.35(4) &  0.73(9)  & 6.44(7) &  1.16(7)& 2.32(1) & 6.54(9) &  0.4(1)  & 0.52(8) &1.70(67) &4.80/1.13  \\
60022-01-01-00 &  51998.64  & 1.06(3) & 2.49(1) &  0.04(1) & 12.03(1) &  1.20(5) & 2.86(2) & 6.24(2) &  0.5(1  & 1.50(6) &1.07(67) &7.94/1.69  \\
60022-01-01-01 &  51998.71  & 1.03(9) & 2.52(2) &  0.07(4) & 12.97(3) &  1.20(7) & 2.31(3) & 6.5(1)  &  0.56(8)& 1.33(5) &1.09(67) &8.43/1.91  \\
60022-01-01-02 &  51998.78  & 1.01(3) & 2.57(1) & -0.04(1) & 11.92(3) &  1.2(1)  & 3.26(2) & 6.59(5) &  0.58(7)& 1.33(7) &0.87(67) &8.43/1.91  \\
60022-01-01-03 &  51999.78  & 1.02(7) & 2.47(3) &  0.04(1) & 12.31(4) &  1.20(3)  & 2.39(4) & 6.55(4) &  0.55(7)& 1.57(6) &1.15(67) & 8.02/1.69 \\
60022-01-02-00 &  52028.61  & 1.00(9) & 2.51(6) &  0.03(1) & 11.50(3) &  1.20(5) & 2.74(4) & 6.56(2) &  0.53(6) & 1.22(5) &1.13(67) & 7.56/1.62 \\
60022-01-02-01 &  52028.94  & 1.01(3) & 2.54(2) &  0.01(1) & 11.73(2) &  1.20(6) & 2.51(6) & 6.50(6) &  0.59(5) & 1.23(3) &1.05(67) & 7.62/1.65 \\
60022-01-01-04 &  52029.47  & 1.07(3) & 2.52(1) &  0.06(2) & 13.88(6) &  1.17(9)  & 3.15(2) & 6.50(7) &  0.62(7)& 1.61(8) &0.98(67) & 9.15/2.02 \\
60022-01-01-05 &  52029.61  & 1.03(6) & 2.51(2) &  0.2(1)  & 13.36(3) &  1.20(5)  & 2.39(6) & 6.51(6) &  0.50(8)& 1.03(5) &0.88(67) & 8.80/2.21 \\
60022-01-01-06 &  52029.69  & 1.05(3) & 2.53(4) & -0.01(1) & 11.54(4) &  1.20(4)  & 2.84(9) & 6.54(2) &  0.53(9)& 1.26(6) &1.40(67) & 7.57/1.58 \\
60022-01-03-00 &  52067.39  & 1.01(4) & 2.53(1) &  0.21(7) & 13.35(3) &  1.20(7)  & 2.62(4) & 6.51(2) &  0.5(1) & 1.22(3) &1.09(67) & 8.87/2.22 \\
60022-01-03-01 &  52067.58  & 1.03(5) & 2.52(3) &  0.04(1) & 12.65(3) &  1.20(4) & 2.36(5) & 6.55(8) &  0.35(4) & 1.28(3) &1.30(67) & 8.22/1.82 \\
60022-01-03-01 &  52068.30  & 1.07(9) & 3.16(2) & -0.46(9) & 10.82(4) &  1.20(5) & 5.41(3) & 6.36(3) &  0.41(8) & 1.58(4) &1.20(67) & 8.24/1.94 \\
60022-01-04-03 &  52102.363 & 1.03(9) & 3.25(9) & -0.57(8) & 9.44(7) &  1.31(7)  & 5.47(4) & 7.07(3) &  0.50(9) & 1.64(5) &1.07(67) & 7.44/1.60 \\
60022-01-04-00 &  52103.361 & 1.01(5) & 2.72(5) & -0.20(8) & 9.7(1)  &  1.30(6)  & 6.65(9) & 7.09(4) &  0.52(9) & 1.86(5) &1.24(67) & 8.38/1.78 \\
60022-01-04-02 &  52103.685 & 1.0(1)  & 2.87(6) & -0.1(1)  & 10.68(9) &  1.20(7) & 4.01(9) & 7.03(3) &  0.50(5) & 1.63(3) &1.04(67) & 7.78/1.72 \\
60022-01-01-07 &  52127.54  & 0.99(3) & 2.53(7) &  0.24(1) & 13.08(3) &  1.18(4) & 2.64(9) & 6.58(3) &  0.36(8) & 1.14(3) &1.10(67) & 8.71/2.21 \\
60022-01-03-03 &  52128.667 & 0.99(5) & 2.78(5) & -0.26(5) & 10.23(1) &  1.20(8) & 6.8(1)  & 6.34(5) &  0.4(1)  & 1.57(3) &1.04(67) & 8.24/1.94 \\
60022-01-04-04 &  52128.734 & 1.0(1) & 2.93(5) & -0.29(8)  & 10.40(7) &  1.20(9) & 5.8(1)  & 7.09(4) &  0.38(7) & 1.65(4) &1.05(67) & 8.04/1.96 \\
60022-01-03-04 &  52129.386 & 1.02(6) & 2.72(6) & -0.17(7) & 9.9(1)  &  1.20(4)  & 5.97(9) & 6.42(3) &  0.37(8) & 1.83(3) &1.01(67)& 8.38/1.86 \\
60022-01-04-01 &  52131.114 & 1.07(6) & 2.87(6) & -0.3(1)  & 9.36(3) &  1.20(8)  & 4.56(6) & 7.09(5) &  0.46(6) & 1.64(4) &1.15(67)& 7.15/1.53 \\
60022-01-05-00 &  52131.507 & 1.0(1) & 2.62(5) & -0.09(1)  & 8.5(1)  &  1.30(6)  & 5.8(1)  & 5.99(7) &  0.47(4) & 1.69(3) &1.16(67)& 7.49/1.60 \\
60022-01-05-01 &  52131.375 & 0.99(7) & 2.89(8) & -0.36(1) & 8.2(1)  &  1.20(4)  & 6.5(1)  & 5.98(7) &  0.41(8) & 1.72(6) &1.05(67)& 7.61/1.64 \\
60022-01-06-01 &  52166.493 & 1.0(1)  & 2.75(3) & -0.17(2) & 11.19(5)&  1.20(9)  & 5.53(5) & 5.87(7) &  0.43(9) & 1.61(3) &0.96(67)& 8.73/2.13 \\
60022-01-06-03 &  52168.754 & 1.00(7) & 2.96(6) & -0.37(5) & 11.36(6)&  1.2(1)   & 5.52(9) & 5.57(8) &  0.39(4) & 1.63(3) &1.16(67)& 8.76/2.05 \\
60022-01-06-05 &  52169.610 & 1.01(4) & 2.65(4) &  0.08(7) & 11.04(7)&  1.20(5)  & 5.26(8) & 6.17(3) &  0.418)  & 1.65(6) &1.06(67)& 8.34/2.31 \\
60022-01-06-06 &  52169.678 & 1.02(5) & 3.04(6) & -0.40(3) & 12.17(8) &  1.20(9) & 5.22(7) & 6.28(2) &  0.38(6) & 1.63(8) &1.11(67)& 8.94/2.20  \\
60022-01-06-07 &  52170.534 & 1.03(5) & 2.6(1) & -0.04(3)  & 7.20(4) &  1.20(5)  & 5.33(3) & 6.47(7) &  0.50(8) & 1.68(7) &1.12(67)& 6.58/1.38  \\
60022-01-06-00 &  52170.983 & 1.00(5) & 2.65(7) & -0.15(5) & 9.4(1) &  1.30(6)   & 6.53(7) & 5.87(6) &  0.39(7) & 1.64(4) &1.1(67)& 7.98/1.72  \\
60022-01-06-09 &  52171.194 & 1.06(7) & 2.86(5) & -0.23(4) & 11.1(1)  &  1.20(7) & 5.70(5) & 6.47(8) &  0.37(9) & 1.74(5) &1.06(67)& 8.75/2.13  \\
60022-01-06-08 &  52172.176 & 1.04(4) & 2.99(8) & -0.43(6) & 10.40(5) &  1.20(9) & 5.39(6) & 6.57(6) &  0.4(1)  & 1.71(4) &1.1(67)& 8.22/1.82  \\
60022-01-06-10 &  52172.241 & 1.0(1)  & 2.66(4) & -0.25(5) & 10.03(4) &  1.20(5) & 6.26(8) & 6.43(7) &  0.40(8) & 1.68(3) &0.95(67)& 8.44/2.02  \\
60022-01-06-11 &  52172.725 & 0.99(6) & 2.62(9) & -0.49(7) & 9.34(5) &  1.30(6)  & 6.34(5) & 6.40(3) &  0.38(7) & 1.74(2) &1.08(67)& 8.30/1.83  \\
60022-01-06-12 &  52172.792 & 1.00(3) & 3.41(9) & -0.54(4) & 12.08(7) &  1.20(8) & 5.24(9) & 6.81(5) &  0.35(8) & 1.81(5) &1.01(67)& 8.84/2.15  \\
60022-01-06-13 &  52172.852 & 1.03(5) & 2.83(3) & -0.20(3) & 11.63(6) &  1.20(5) & 6.09(9) & 6.17(7) &  0.32(5) & 1.77(7) &1.05(67)& 8.88/2.24  \\
60022-01-07-00 &  52201.445 & 1.04(4) & 2.75(5) & -0.13(3) & 10.13(3) &  1.20(6) & 5.50(8) & 6.24(8) &  0.35(8) & 1.69(3) &1.07(67)& 8.06/1.98  \\
60022-01-07-02 &  52204.424 & 1.06(9) & 2.75(5) & -0.23(6) & 9.11(4) &  1.20(5)  & 5.42(9) & 6.36(6) &  0.3(1)  & 1.65(4) &0.89(67)& 7.57/1.67  \\
60022-01-07-03 &  52207.202 & 1.03(5) & 2.76(3) & -0.02(1) & 11.05(5) &  1.20(7) & 5.07(9) & 5.85(7) &  0.35(8) & 1.62(3) &1.15(67)& 8.31/2.32  \\
60022-01-07-04 &  52207.071 & 1.01(4) & 2.69(3) & -0.01(1) & 10.37(3) &  1.20(3) & 5.56(8) & 6.52(5) &  0.4(1)  & 1.76(5) &1.12(67)& 8.26/2.06  \\
60022-01-08-00 &  52230.164 & 1.0(5) & 2.58(4) & -0.01(1)  &7.64(1)   &  1.3(1)  & 5.67(7) & 5.87(7) &  0.50(9) &  1.64(3)& 1.15(67)& 6.84/1.49 \\
60022-01-09-00 &  52295.306 & 1.0(4) & 2.51(4) &  0.23(1)  &12.25(4) &  1.2(1)  & 3.53(8) & 5.97(8)  &  0.46(8) &  1.67(6)& 1.14(67)& 8.55/2.04 \\
60022-01-10-00 &  52321.559 & 1.0(5) & 2.50(4) &  0.11(1)  &13.26(4) &  1.20(9) & 2.71(8) & 5.85(9)  &  0.41(6) &  1.71(8) & 1.2(67)&  8.72/1.97 \\
60022-01-10-01 &  52320.301 & 1.0(6) & 2.51(2) &  0.09(1)  &13.77(1) &  1.20(6) & 2.61(8) & 5.78(7)  &  0.46(7) &  1.61(4) & 1.1(67)&  9.08/2.18 \\
60022-01-10-02 &  52320.876 & 1.0(3) & 2.52(1) &  0.06(1)  &12.19(4) &  1.20(5) & 2.60(8) & 5.87(6)  &  0.4(1)  &  1.65(4) & 1.04(67)& 8.01/1.76 \\
60022-01-10-03 &  52324.647 & 0.99(2) & 2.44(2) &  0.14(5) & 11.36(4)&  1.10(8) & 1.57(6) & 6.26(6) &  0.47(9)  & 1.87(3) &1.11(67)& 6.39/1.37  \\
60022-01-10-04 &  52325.760 & 0.99(2) & 2.44(3) &  0.15(6) & 11.91(5)&  1.10(9) & 2.23(6) & 6.47(7) &  0.51(6)  & 1.79(6) &1.09(67)& 6.39/1.37  \\
60022-01-08-00 &  52230.164 & 1.00(5) & 2.58(4) & -0.02(1) & 7.64(2) &  1.2(1)  & 5.67(7) & 6.75(6) &  0.53(8)  & 1.86(4) &0.88(67)& 6.84/1.49  \\
60022-01-11-00 &  52356.747 & 0.99(3) & 2.50(2) &  0.08(7) & 9.94(6) &  1.10(8) & 1.29(6) & 6.55(4) &  0.49(6)  & 1.86(7) &1.2(67)& 6.39/1.37  \\
60022-01-11-01 &  52356.814 & 0.99(1) & 2.47(1) &  0.12(4) & 11.79(3)&  1.10(5) & 1.58(6) & 6.46(3) &  0.5(1)   & 1.85(7) &1.15(67)& 7.54/1.64  \\
60022-01-11-03 &  52357.083 & 0.99(2) & 2.41(1) &  0.6(1)  & 13.29(4)&  1.10(8) & 2.8(1)  & 6.57(7) &  0.50(9)  & 1.87(4) &1.1(67)& 8.56/1.94  \\
60022-01-11-02 &  52394.572 & 1.0(1)  & 2.51(2) &  0.41(4) & 12.56(4)&  1.10(4) & 6.2(1)  & 6.38(5) &  0.39(7)  & 1.94(3) &1.19(67)& 9.92/2.36  \\
60022-01-12-00 &  52394.891 & 1.0(1)  & 2.73(4) &  0.08(7) & 13.26(7)&  1.10(7) & 5.76(8) & 6.49(9) &  0.37(8)  & 1.85(6) &1.01(67)& 9.92/2.36  \\
60022-01-13-00 &  52554.152 & 1.0(1)  & 2.46(2) &  0.6(1)  & 12.5(9) &  1.1(1)  & 9.48(4) & 6.35(5) &  0.35(4)  & 1.91(4) &1.07(67)& 9.92/2.36  \\
60022-01-13-01 &  52554.484 & 1.00(5) & 2.56(5) &  0.03(1) & 13.6(7) &  1.20(5) & 2.85(7) & 6.48(7) &  0.36(9)  & 1.86(3) &1.16(67)& 8.87/1.99  \\
90022-05-01-00 &  53108.101 & 1.04(8) & 2.42(9) & -0.18(6) & 4.22(3) &  1.5(1)  & 4.07(3) & 6.38(6) &  0.53(2)  & 0.87(5) &1.36(67)& 3.85/0.79  \\
90022-05-01-01 &  53108.256 & 1.00(8) & 3.1(1)  & -0.46(7) & 4.75(7) &  1.59(9) & 3.68(5) & 6.34(5) &  0.60(8)  & 1.03(7) &1.38(67)& 3.99/0.84   \\
90022-05-02-00 &  53238.243 & 1.01(8) & 2.56(4) & 0.10(6)  & 4.20(7) &  1.19(7) & 2.58(9) & 6.39(9) &  0.62(3)  & 0.77(9) &1.34(67)& 3.07/0.64   \\
90022-05-03-00 &  53245.990 & 1.06(9) & 2.36(2) & 0.30(6) & 5.42(5) &  1.19(8)  & 1.5(1)  & 6.5(1)  &  0.6(1)   & 0.79(8) &1.00(67)& 3.76/0.75   \\
90022-05-04-00 &  53252.490 & 1.00(4) & 2.5(1)  & 0.10(7) & 6.05(9) &  1.19(6)  & 1.51(8) & 6.6(1)  &  0.65(8)  & 0.86(9) &1.00(67)& 3.11/0.83   \\
90022-05-04-01 &  53253.859 & 1.0(1)  & 2.47(2) & 0.06(2) & 6.08(3) &  1.50(6)  & 2.49(5) & 6.50(9) &  0.66(6)  & 0.83(5) &1.25(67)& 4.23/0.85    \\
90022-05-05-00 &  53263.468 & 1.01(4) & 2.52(1) & 0.04(1) & 5.33(2) &  1.50(8)  & 2.24(2) & 6.49(7) &  0.67(8)  & 0.97(7) &1.15(67)& 4.23/0.85    \\
90022-05-06-00 &  53265.412 & 1.09(6) & 2.52(2) & 0.01(1) & 5.42(3) &  1.45(7)  & 2.46(3) & 6.6(1)  &  0.63(6)  & 0.87(5) &1.15(67)& 3.78/0.75    \\
90022-05-06-01 &  53265.541 & 1.0(1)  & 2.50(1) & 0.01(1) & 5.56(5) &  1.39(5)  & 2.08(4) & 6.62(9) &  0.62(4)  & 0.79(7) &1.05(67)& 3.79/0.77    \\
90022-05-06-02 &  53265.675 & 1.01(2) & 2.51(2) & 0.04(2) & 5.54(6) &  1.41(4)  & 2.34(6) & 6.67(5) &  0.61(8)  & 0.79(5) &1.07(67)& 3.82/0.78    \\
90022-05-06-03 &  53265.806 & 1.02(5) & 2.56(2) & -0.02(1) & 4.98(3) &  1.6(1)  & 2.22(5) & 6.6(1)  &  0.62(9)  & 0.81(9) &1.08(67)& 3.44/0.69    \\
90022-05-07-00 &  53279.124 & 1.00(7) & 2.53(1) & 0.10(6) & 3.53(1) &  1.70(8)  & 2.45(3) & 6.64(7) &  0.64(5)  & 0.85(6) &1.18(67)& 3.44/0.69    \\
90022-05-07-00 &  53280.960 & 1.03(9) & 2.42(1) & 0.31(3) & 5.42(3) &  1.50(9)  & 2.18(3) & 6.65(8) &  0.62(8)  & 0.79(7) &0.97(67)& 3.44/0.69    \\
94307-05-01-00 &  55129.659 & 1.02(3) & 3.16(4) & 0.41(3) & 4.53(7) &  1.4(1)   & 3.08(2) & 6.58(2) &  0.4(1)   & 0.32(1) &1.09(67)& 3.21/1.28    \\
94307-05-01-000&  55440.307 & 1.04(4) & 3.16(5) & 0.45(4) & 4.53(7) &  1.50(8)  & 3.08(3) & 6.59(5) &  0.40(7)  & 0.34(2) &0.87(67)& 3.21/1.28    \\
94307-05-01-00 &  55440.62  & 1.03(8) & 2.50(2) & 0.25(2) & 4.67(2) &  1.4(1)   & 4.17(4) & 6.64(9) &  0.62(9)  & 0.68(9) &1.13(67)& 3.78/0.78    \\
94307-05-01-01 &  55440.701 & 1.03(9) & 3.16(4) & 0.41(3) & 4.53(7) &  1.40(6)  & 3.08(2) & 6.58(3) &  0.4(1)   & 0.32(6) &0.89(67)& 3.21/1.28    \\
94307-05-01-02 &  55440.766 & 1.02(3) & 3.26(7) & -0.56(3) & 4.57(1) &  1.35(7) & 4.13(2) & 6.21(2) &  0.5(1)   & 0.21(7) &0.98(67)& 3.21/1.28    \\
      \enddata%     \hline%      \end{tabular}
    \label{tab:fit_table_rxte}
%  \end{center}
$^\dagger$ The spectral model is  $wabs*(blackbody + COMPTB + Gaussian)$, where $N_H$ is fixed at 
a value 1.6$\times 10^{22}$ cm$^{-2}$ (Oosterbroek et al., 2000); 
color temperature %$T_s$ and 
$T_{BB}$ is fixed at %1.3 and 
0.6 keV %, respectively
 (see comments in the text); 
%$^{\dagger\dagger}$ when parameter $\log(A)\gg1$, this parameter is fixed at 1.0 (see comments in the text), 
$^{\dagger\dagger}$ normalization parameters of $blackbody$ and $COMPTB$ components are in units of 
$L_{37}^{soft}/d^2_{10}$, where $L_{37}^{soft}$ is the source luminosity in units of 10$^{37}$ erg/s, 
$d_{10}$ is the distance to the source in units of 10 kpc 
and $Gaussian$ component is in units of $10^{-2}\times total~~photons$ $cm^{-2}s^{-1}$ in line 
%({\bf Lenochka please specify units for these components, look at XSPEC manual})
,  
%$\sigma_{line2}$ of Gaussian2 component is fixed to a value 0.01 keV (see comments in the text), 
$^{\dagger\dagger\dagger}$spectral fluxes (F$_1$/F$_2$) in units of $\times 10^{-9}$ ergs/s/cm$^2$ for  (3 -- 10) and (10 -- 50) keV energy ranges respectively.  
%in units of $\times 10^{-9}$ ergs/s/cm$^2$.
%* this observations are  fitted with $bmc+Gaussian1+Gaussian2+bbody$ model, see values of the best-fit BB color temperature and EW in Table 2, 3 and 4.
\end{deluxetable}
%~~~~~~~~~~
\vspace{2.in}
\newpage
~~~~~~~~
%\vspace{2.in}
\begin{deluxetable}{llcccccccc}
%\begin{deluxetable}{cccccccccccccccc}
\rotate
\tablewidth{0in}
\tabletypesize{\scriptsize}
%  \begin{center}
    \tablecaption{Comparisons of the best-fit parameters  of ÒatollÓ sources GX~3+1 and 4U~1728-34$^1$}
%\vspace{1em}
    \renewcommand{\arraystretch}{1.2}
%    \begin{tabular}[h]
%      \hline
%ID               & day  & & &   $\Gamma-1$          &           &$L_{39}/d^2_{10}$& keV & keV   &  &  keV        &  & & & }
 \tablehead
{Source & Alternative & Class$^2$& Distance, & Presence of & $kT_e$,  & $ N_{COM}$ &$kT_{BB},$ &  $kT_{s},$  & $f$ \\
  name  & name        &          & kpc       & kHz QPO     & keV         &  $L_{39}^{soft}/{D^2_{10}}$        & keV  & keV  &  }
 \startdata%   id     MJD      kT_Bbody     N_bb     [kT_s]    alf       T_e       log    norm_COMPTB E_line[Sigma_l] N_line Xi_2(dof)  Flux3-10 Fl10-60
4U~1744-26  & GX 3+1   & Atoll, Sp, B & 4.5$^3$ &     none$^5$    & 2.3-4.5 &  0.04-0.15 & 0.6 & 1.16-1.7 & 0.2-0.9   \\
4U 1728-34  & GX~354-0 & Atoll, Su, D & 4.2-6.4$^4$ & +$^6$&  2.5-15& 0.02-0.09 & 0.6-0.7 & 1.3 & 0.5-1\\     
      \enddata%     \hline
%      \end{tabular}
    \label{tab:fit_table_comb}
%  \end{center}
%$^\dagger$ The spectral model is  $wabs*(blackbody + COMPTB + Gaussian)$,
%normalization parameters of $blackbody$ and $COMPTB$ components are in units of 
%$L_{37}^{soft}/d^2_{10}$ 
%$erg/s/kpc^2$, 
%where $L_{37}^{soft}$ is the soft photon  luminosity in units of 10$^{37}$ erg/s, 
%$d_{10}$ is the distance to the source in units of 10 kpc 
%and $Gaussian$ component is in units of $10^{-2}\times total~~photons$ $cm^{-2}s^{-1}$ in line.
%$^{\dagger\dagger\dagger}$when parameter $\log(A)\gg1$, it is fixed to a value 1.0 (see comments in the text). 
%$\sigma_{line2}$ of Gaussian2 component is fixed to a value 0.01 keV (see comments in the text), 
%$^{\dagger\dagger\dagger\dagger}$spectral fluxes (F$_1$/F$_2$) in the (3 -- 10)/(10 -- 60) keV energy ranges, correspondingly, 
%in units of $\times 10^{-9}$ ergs/s/cm$^2$.
References:
(1) ST11,
(2) Classification of the system in the varios schemes (see text): Sp = supercritical, Su = subcritical, 
B = bulge, D = disk,
(3) \citet{kk00},
(4) \citet{par78},
(5) \citet{stroh98}
(6) \citet{to99}
\end{deluxetable}

\newpage

\vspace{10.in}

\newpage

\newpage 

\newpage

\newpage

\newpage

\newpage

%\newpage
%\begin{figure}[ptbptbptb]
%\includegraphics[scale=0.9,angle=0]{chi2_4.eps}
%\caption{Function of 
%$\chi^2(\Gamma_{mean})=\frac{1}{N}sum_{i=1}^N(\frac{\Gamma_i-\Gamma_{mean}}{\Delta\Gamma_i})^2$ 
%versus $\Gamma_{mean}$. The dot-dashed and dashed horizontal lines indicate the critical residuals $\chi^2_{red}$ for the  
%1\% and 33\% (1$\sigma$) significance levels. The {\it bright blue, pink} and $green$ curves correspond to fits with 0.5\%, 0.7\% and 1\% 
%of systematic errors.
%}
%\label{chi2}
%\end{figure}

\newpage

\newpage

\newpage

\newpage

%\newpage
%\begin{figure}[ptbptbptb]
%%%\includegraphics[scale=0.9,angle=0]{cmpt_gx_4U_1.ps}
%\includegraphics[scale=0.9,angle=0]{f12.eps}
%\caption{Similar to  that presented in Fig. \ref{lc_1998} but  for all the {\it RXTE} 
%2000 outburst transition 
%sets ({\it R1 - R7}) for illustradion of $slow~(long-term)$ variability.
%}
%\label{evolution cmpt vs T_e_1}
%\end{figure}

%\newpage
%\begin{figure}[ptbptbptb]
%\includegraphics[scale=0.9,angle=0]{compt_gx_4U_2.ps}
%%%%%%%\includegraphics[scale=0.9,angle=0]{f11a.eps}
%\caption{Similar to  that presented in Fig. \ref{lc_1998} but  for all the {\it RXTE} 
%%2000 outburst transition 
%sets ({\it R1 - R7}) for illustradion of $slow~(long-term)$ variability.
%}
%\label{evolution cmpt vs T_e_2}
%\end{figure}

\end{document}